\documentclass[twocolumn,pdflatex,sn-nature]{sn-jnl}
\usepackage{multirow}%
\usepackage{amsmath,amssymb,amsfonts}%
\usepackage{amsthm}%
\usepackage{mathrsfs}%
\usepackage[title]{appendix}%
\usepackage{xcolor}%
\usepackage{textcomp}%
\usepackage{manyfoot}%
\usepackage{graphicx}   %
\usepackage{subcaption}
\usepackage{svg}
\date{} 

\begin{document}

\title[Machine Learned Potential for High-Throughput Phonon Calculations of Metal--Organic Framework]{Machine Learned Potential for High-Throughput Phonon Calculations of Metal--Organic Frameworks}
\author[1]{\fnm{Alin Marin} \sur{Elena}}
\equalcont{These authors, ordered alphabetically, contributed equally. All others, except for the corresponding author, are also ordered alphabetically.}

\author[2,3]{\fnm{Prathami Divakar} \sur{Kamath}}
\equalcont{These authors, ordered alphabetically, contributed equally. All others, except for the corresponding author{\text{\color{blue}*}}, are also ordered alphabetically.}

\author[2,3,4]{\fnm{Théo} \sur{Jaffrelot Inizan}}

\author[5]{\fnm{Andrew S.} \sur{Rosen}}

\author[1]{\fnm{Federica} \sur{Zanca}}

\author*[2,3]{\fnm{Kristin A.} \sur{Persson}}\email{kapersson@lbl.gov}

\affil[1]{\orgdiv{Scientific Computing Department}, \orgname{Science and Technology Facilities Council},
\orgaddress{\street{Keckwick Lane}, \city{Daresbury}, \postcode{WA4 1PT}, \state{Cheshire}, \country{UK}}}

\affil[2]{\orgdiv{Department of Materials Science and Engineering}, \orgname{University of California}, \orgaddress{\city{Berkeley}, \postcode{CA 94720},
\state{California}, \country{USA}}}

\affil[3]{\orgdiv{Materials Sciences Division}, \orgname{Lawrence Berkeley National Laboratory}, \orgaddress{ \city{Berkeley},
\postcode{CA 94720}, \state{California},
\country{USA}}}

\affil[4]{\orgdiv{Bakar Institute of Digital Materials for the Planet, College of Computing, Data Science, and Society}, \orgname{University of California},  \orgaddress{ \city{Berkeley},
\postcode{CA 94720}, \state{California},
\country{USA}}}

\affil[5]{\orgdiv{Department of Chemical and Biological Engineering}, \orgname{Princeton University}, \orgaddress{ \city{Princeton},
\postcode{NJ 08544}, \state{New Jersey},
\country{USA}}}

 \abstract{
         Metal--organic frameworks (MOFs) are highly porous and versatile materials studied extensively for applications such as carbon capture and water harvesting. However, computing phonon-mediated properties in MOFs, like thermal expansion and mechanical stability, remains challenging due to the large number of atoms per unit cell, making traditional Density Functional Theory (DFT) methods impractical for high-throughput screening. Recent advances in machine learning potentials have led to foundation atomistic models, such as MACE-MP-0, that accurately predict equilibrium structures but struggle with phonon properties of MOFs. In this work, we developed a workflow for computing phonons in MOFs within the quasi-harmonic approximation with a fine-tuned MACE model, MACE-MP-MOF0. The model was trained on a curated dataset of 127 representative and diverse MOFs. The fine-tuned MACE-MP-MOF0 improves the accuracy of phonon density of states and corrects the imaginary phonon modes of MACE-MP-0, enabling high-throughput phonon calculations with state-of-the-art precision. The model successfully predicts thermal expansion and bulk moduli in agreement with DFT and experimental data for several well-known MOFs. These results highlight the potential of MACE-MP-MOF0 in guiding MOF design for applications in energy storage and thermoelectrics.}
\maketitle

\section{Introduction}
Metal--organic frameworks (MOFs) are composed of organic molecules, called linkers, connected to inorganic ions or clusters, called nodes \cite{mof1,mof2,mof3}. MOFs are nanoporous materials that, due to their large modular nature and high porosity, have become strong candidates for many potential applications ranging from water harvesting \cite{water, water2} and catalysis to biosensing. Some important physical properties that make MOFs suitable for such potential applications include mechanical stability, thermal expansion, heat conduction and superconductivity which are influenced by phonon-mediated lattice dynamics \cite{D3TA02214E,thermal,deform,supercond}. Phonons describe the collective vibrations of atoms in a crystal and interact with electronic and thermal excitations that affect these physical characteristics of MOFs \cite{phonons}. 
However, the current understanding of these properties in MOFs is limited due to the computational complexity of predicting phonons in these large structures. Currently, one of the most reliable methods to study the electronic structure of materials and their properties is Density Functional Theory (DFT) \cite{PhysRev.136.B864,dft,dreizler_gross_dft_1990}. However, for materials like MOFs that can have several hundreds or even thousands of atoms in their unit cell, performing supercell calculations necessary for obtaining accurate phonons becomes too computationally expensive for screening applications. 

Semi-empirical Quantum Mechanical methods, such as the Density Functional Tight Binding (DFTB) method, have been used to compute phonons in few highly symmetric isoreticular MOFs \cite{PhysRevMaterials.3.116003}. Several DFTB schemes have been developed, with DFTB3 \cite{dftb3} being the most widely used. While DFTB is 2–3 orders of magnitude faster than DFT, its pair- wise parametrization strategy and lack of available metal atom parameters limits its use in screening applications, particularly pertaining to phonons in MOFs. A notable variant of DFTB3, GFN1-xTB \cite{tb} follows a element-specific parameters strategy and extends its applicability to nearly the entire periodic table, up to thousands of atoms. In \cite{D2CP00184E}, the GFN1-xTB method has been reported to produce cell parameters within a 5\% deviation (relative to experiments) for 75\% of the MOFs in the CoRE database,\cite{core1,core2} showing promise for adsorption applications and obtaining binding energies. However, GFN1-xTB still scales quadratically with the number of electrons which may present challenges for larger MOFS, and has not been used for the analysis of vibrational properties in MOFs. 

Traditional force fields like UFF \cite{uff} and CHARMM \cite{Brooks2009} are widely used due to their scalability and significant speed up in calculations compared to tight-binding DFT. UFF4MOF \cite{uff4mof}, the UFF parametrization on MOFs, can be used for rapid structure prediction and screening. However, this transferability comes with limitations in accurately predicting dynamical properties such as phonons. \cite{Wieser2024,uff4mof}. Other force fields, like the MOF-FF model \cite{mofff} also predicted accurately the lattice parameters of some well known MOFs like MOF-5, UiO-66 within 0.5\% deviation relative to DFT. Force fields derived with a focus on vibrational properties for MOFs have also been reported like VMOF \cite{C6CP05106E}. While the VMOF model reproduced optimized lattice parameters and phonon density of states in good agreement with DFT, important phonon-derived properties like bulk modulus obtained with VMOF underestimated DFT predictions by more than 50\% even for standard MOFs like UiO-66 and MOF-5 \cite{C6CP05106E}. These limitations of such classical force fields are because of the vast combinations of possible frameworks in the MOF chemical space that make their parameter optimization and selection of functional form extremely challenging.

Recently, several neural network-based machine learning potentials (MLPs) such as \cite{nn,Vandenhaute2023} and on-the-fly MLPs such as kernel-based potentials in the Vienna Ab initio Simulation Package \cite{vasp,vasp1,vasp2,vasp3} (VASP MLPs) and Moment Tensor Potentials (MTP) \cite{mtp} reported in \cite{Wieser2024} have emerged that produce vibrational properties with sufficient accuracy relative to DFT for MOFs. However, such on-the-fly MLPs are restricted to the specific MOF included in the training set, and the need for continuous regeneration and retraining on DFT data to incorporate new configurations renders them impractical for high-throughput screening of MOFs.
Therefore, the need for a ready-to-use transferrable model for screening dynamical properties in MOFs motivates the development of new MLPs.

MLPs like the MACE foundation model (MACE-MP-0) \cite{batatia2024foundation}, which utilizes the MACE architecture \cite{macearc} of an equivariant message-passing graph tensor network with many-body information of atomic features encoded in each layer, has been tested on MOFs. Like several MLPs and force fields that are trained on MOF building blocks \cite{nn, C6CP05106E, uff4mof} rather than on whole MOFs because of their large chemical space and size of the unit cell, the MACE-MP-0 model was trained on the MPtrj dataset of 150k inorganic crystals \cite{mptrj}. The MACE-MP-0 demonstrated high accuracy in predicting the potential energy surface, with a root-mean-squared-error of 33 meV/atom in energies relative to DFT for the 20k MOFs in the QMOF database \cite{MOF,rosen2021machine}. The foundation model thus, demonstrates its transferability in being able to capture some of the complex interactions in MOFs and shows potential to further improve and investigate its accuracy for phonons and derived properties in MOFs. 

In this work, we introduce MACE-MP-MOF0, a highly accurate model derived from MACE-MP-0b (medium model), fine-tuned on a high-quality dataset of 127 representative MOFs. The MACE-MP-0b model is a slightly modified version of the original MACE-MP-0 (which was released in April 2024) to address few shortcomings of the original model such as dealing with short-distance collapse by adding a ZBL potential\footnote{at the time of writing, a new version MACE-MP-0c was released}.The model is evaluated for key properties derived from phonons in well-known and representative MOFs, such as bulk moduli and thermal expansions, within the quasi-harmonic approximation, allowing for a comprehensive comparison with previous models and experimental data reported in the literature. MACE-MP-MOF0 accurately reproduces experimentally observable phenomena, such as negative thermal expansion in MOFs, demonstrating its applicability beyond computational predictions. In addition to analyzing the performance of the model on MOFs in the curated dataset, we test its transferability on other well-known MOFs not seen by the model during the training and find excellent agreement in the predicted bulk modulus wih DFT and experiments. The presented model is therefore ready-to-use for high-throughput phonon calculations of MOFs and aims to incorporate more chemically diverse MOFs in future generations of the model. 

\begin{figure*}[t]
    
    \begin{subfigure}[t]{0.50\textwidth}
     \includegraphics[width=\textwidth]{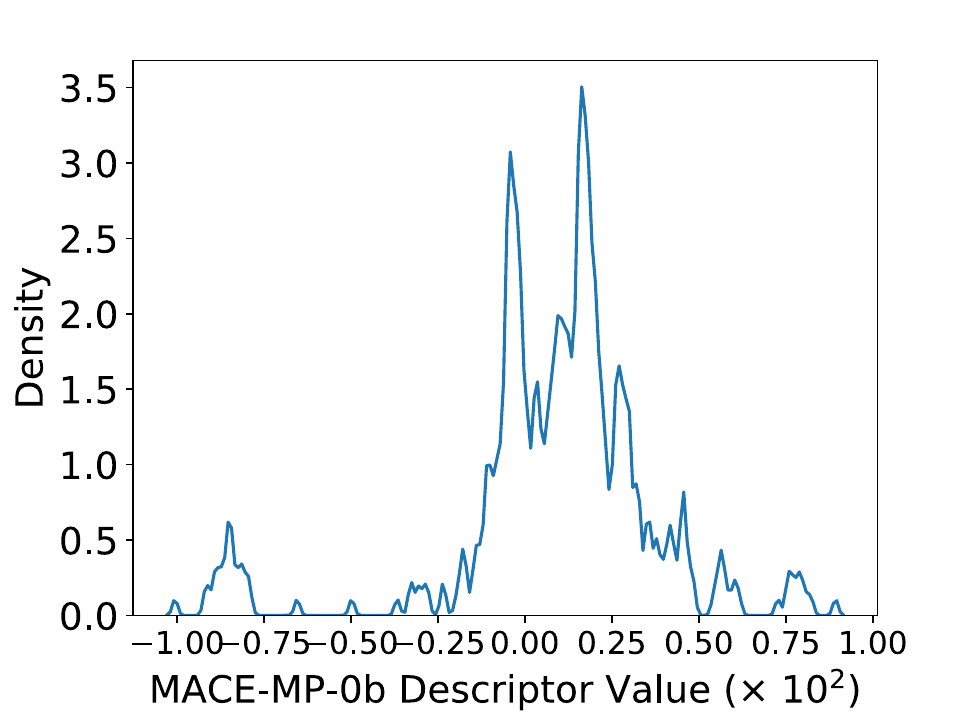}
        \caption*{\textbf{(a)}}
        \label{fig:subfig1}
    \end{subfigure}
    \hfill
    \begin{subfigure}[t]{0.50\textwidth}
        
        \includegraphics[width=\textwidth]{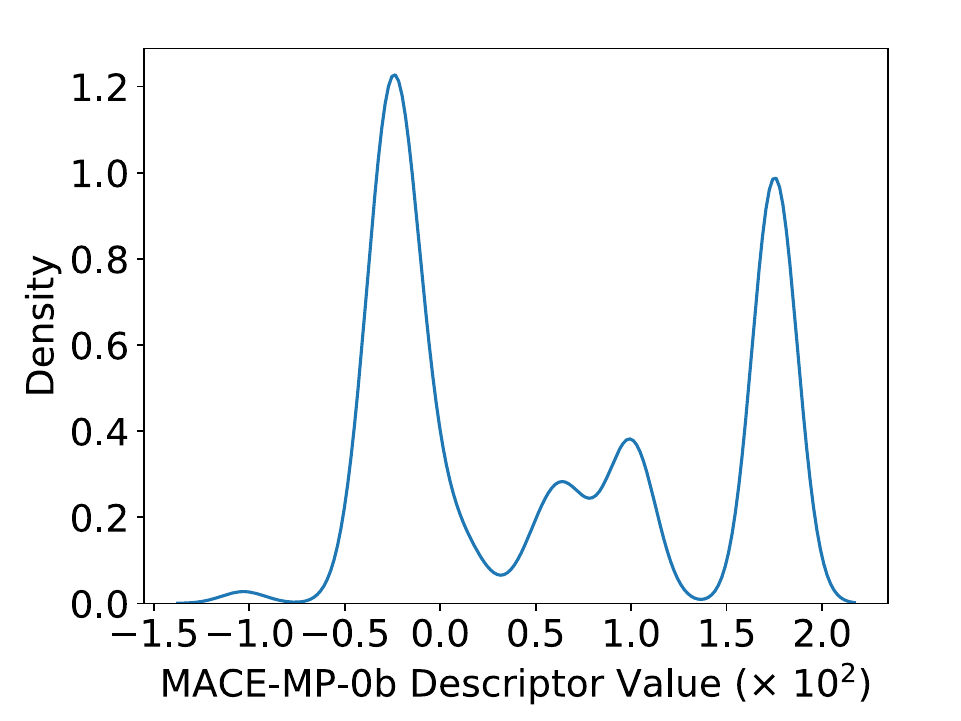}
        \caption*{\textbf{(b)}}
        \label{fig:subfig2}
    \end{subfigure}
    \hfill
    \begin{subfigure}[t]{0.50\textwidth}
        
        \includegraphics[width=\textwidth]{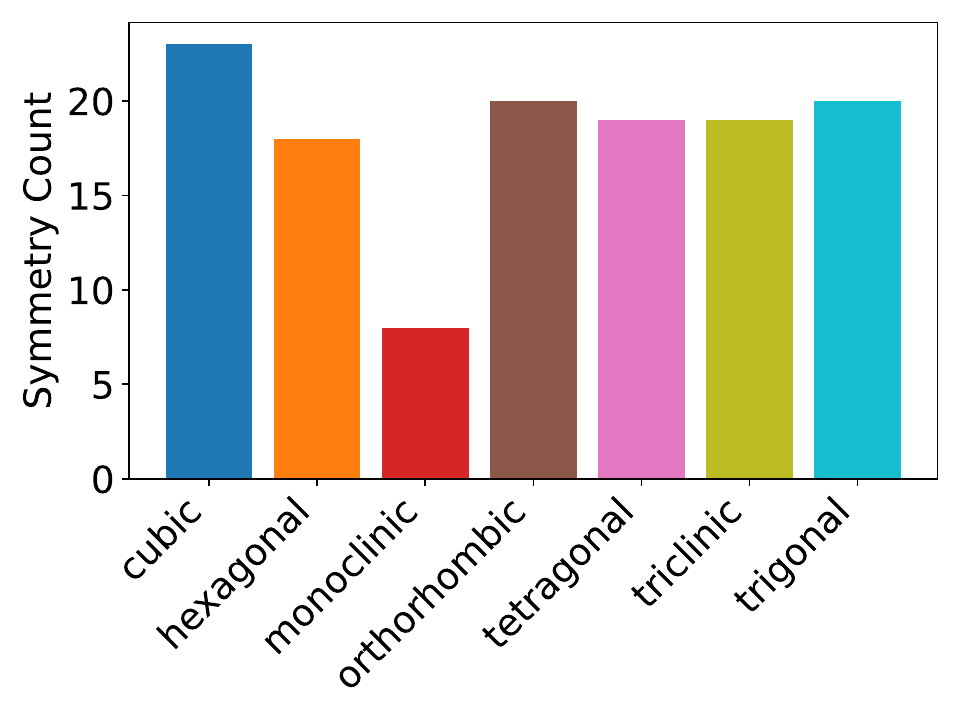}
        \caption*{\textbf{(c)}}
        \label{fig:subfig3}
    \end{subfigure}
    \hfill
    \begin{subfigure}[t]{0.50\textwidth}
        
        \includegraphics[width=\textwidth]{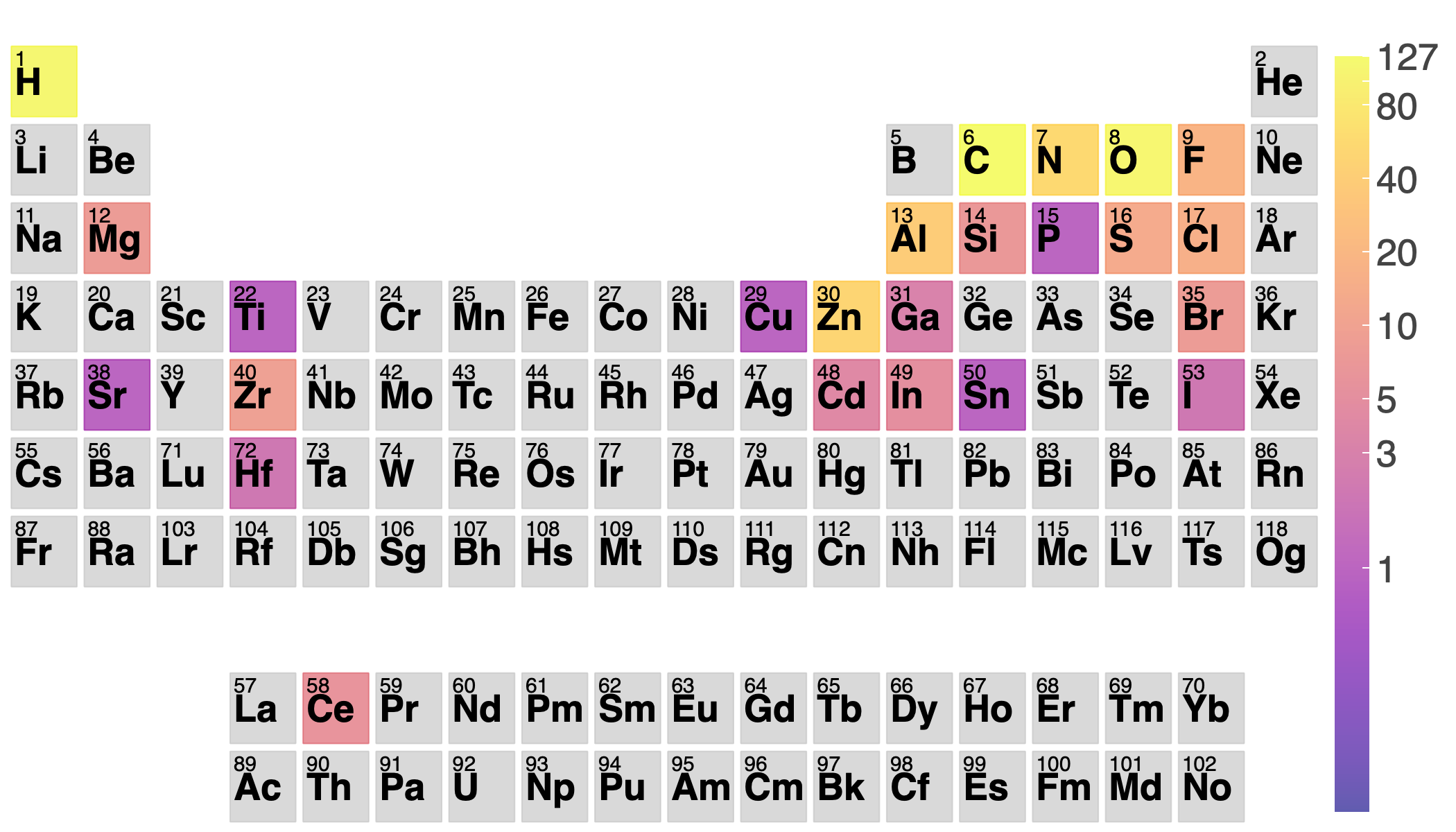}
        \caption*{\textbf{(d)}}
        \label{fig:subfig4}
    \end{subfigure}
    
    \caption{Diversity in the MACE descriptors of the (a) linkers and (b) nodes of the curated 127 MOFs using gaussian kernel density estimation method (c) The symmetry distribution of the dataset (d) The heat map for elemental counts for the atoms included in the dataset. The blank elements in the heat map are not included in the dataset }
    \label{fig: dataset}
    
\end{figure*}

\section{Results}

\subsection{Dataset Curation}
To date, tens of thousands of MOFs have been synthesized \cite{csd}, and countless more can be designed \cite{pormaker} due to the vast diversity of their chemical building blocks, making the selection of an appropriate training set particularly challenging.  In this work, in addition to studying 19 prototypical and experimentally widely studied MOFs for gas storage, catalysis and defect engineering (see Supplementary Information (SI)), we expand our dataset by curating 108 more structures from the 20,375 diverse MOFs in the QMOF database\cite{MOF,rosen2021machine}.

For sampling these 108 structures, we considered predominantly non-spin polarized MOFs with pore limiting diameters (PLDs) greater than 3.6 \AA, using nitrogen as a probe molecule. MACE \cite{macearc} descriptors hold a large amount of information regarding the atomic features of MOFs and hence were used to sample 100 diverse MOFs. These MOFs were selected to span over all the 7 crystal symmetry systems and diverse bonding interactions. Therefore, as shown in Figure \ref{fig: dataset} (c, d), this dataset consists of 127 representative MOFs spread over a wide range of 24 elements in the inorganic clusters and organic ligands. Figure \ref{fig: dataset} (a, b) shows the MACE descriptor space diversity for the curated 127 MOFs (refer the SI for the main features of the sampled MOFs).

On this curated dataset of 127 MOFs, we have used various strategies to generate DFT data points for fine tuning: i) molecular dynamics simulations (an NPT ensemble using MACE-MP-0b+D3), from which frames were selected with a farthest point sampling (FPS) approach to maximize spread in descriptor space (refer SI), ii) strained configurations generated by expanding and compressing unit cells using an equation of state approach, iii) geometry optimization trajectories of various structures by retaining a relatively small number of frames (up to 10, if available), using the FPS approach (refer the Methods section and SI for further details). The aforementioned calculations result in a total of 4764 DFT data points split into 85\% for training and 7.5\% each for testing and validation. We have fine-tuned two models, MACE-MP-MOF0 and MACE-MP-MOF0-v2 where the two models differ only via the way in which the data was split into train, test and validation sets, with the former using a random approach, while the latter using an FPS approach. The goal of comparing the two versions of the fine-tuned model is to show that the performance of these models is comparable regardless of the reference configurations in the training set sampled from this curated dataset.
\vspace{-0.2cm}
\subsection{Phonon Workflow}
\begin{figure*}[t]
    \centering
    \includegraphics[width=\textwidth, height=0.45\textheight]{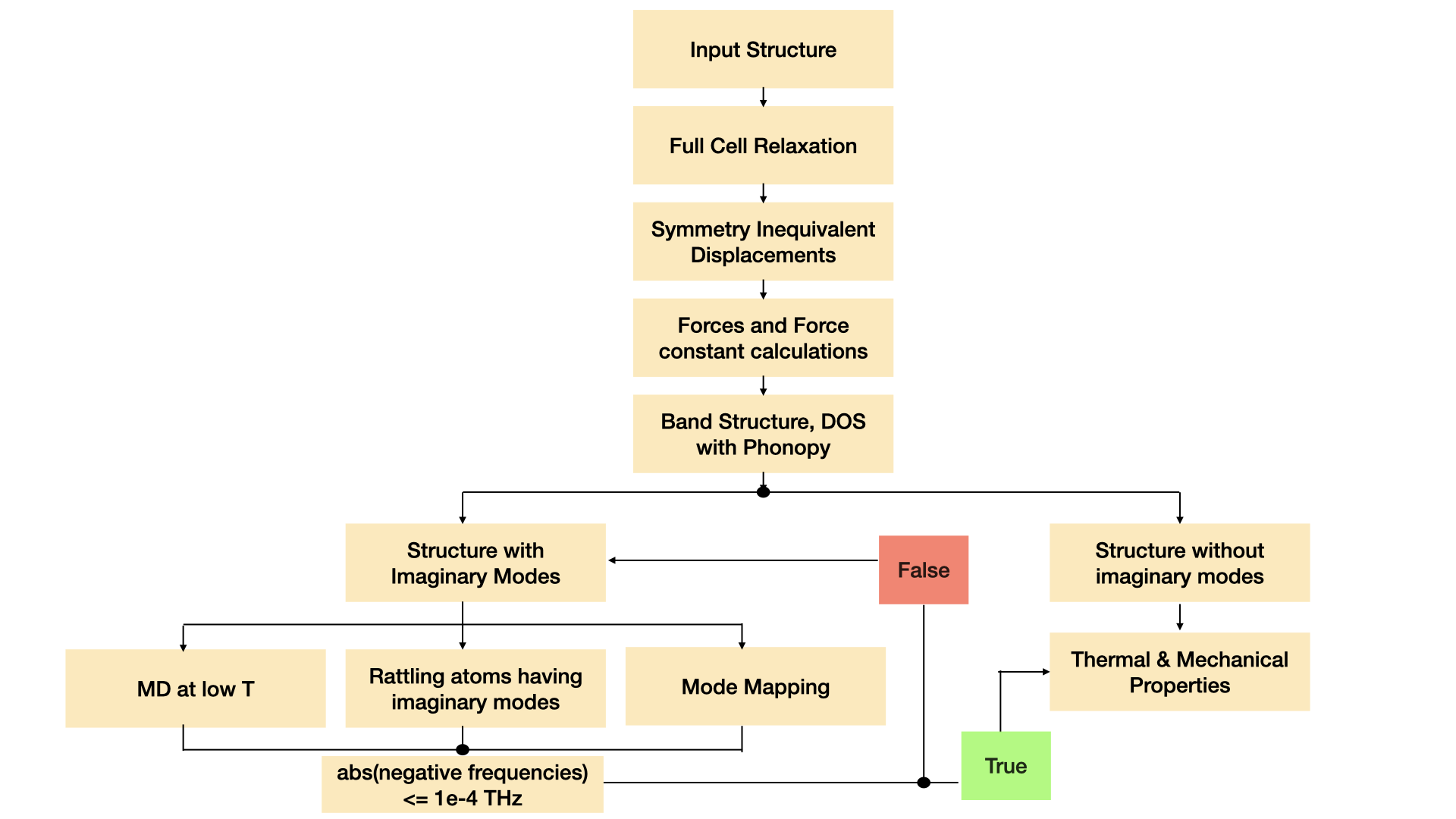}
    \caption{The workflow of MACE-MP-MOF0 for obtaining phonons}
    \label{fig: workflow}
\end{figure*}

As illustrated in Figure \ref{fig: workflow}, the phonon workflow for MACE-MP-MOF0 begins with a full cell relaxation, unconstrained by the symmetry of the input configuration. This step is crucial for ensuring the model's applicability in screening scenarios, particularly when working with MOFs whose stable symmetry configurations are unknown. The full cell relaxation is performed with ASE's \cite{ase-paper} L-BFGS and FrechetCellFilter optimizers until the max force component is $\leq 10^{-6}$ eV/\AA. The search for equilibrium structure is stopped if any negative phonon frequencies present $\leq$ $\left | 10^{-4} \right | $ THz.  The process for eliminating spurious imaginary modes that are larger than this threshold is discussed in the Methods section. The symmetry of the equilibrium structure is determined using Pymatgen's \cite{pymatgen} space group analyzer with a symmetry search tolerance of $10^{-5}$ \text{\AA } for determining number of symmetry inequivalent displacements. 

This work aims to quantitatively compare MACE-MP-MOF0 with models based on various theoretical levels reported in the literature and to ensure consistency across all levels of theory, the Finite Difference (FD) approach is chosen for phonon calculations. The displacements of 0.001 \text{\AA } are produced on the atoms in the $2 \times 2 \times 2$ supercell for obtaining force constants with the FD approach. 

The Harmonic Approximation (HA) for computing phonons fails to capture essential physical properties for MOFs such as thermal expansion, phase transition and elastic moduli \cite{harmonic}. On the other hand, the Quasi-Harmonic approximation (QHA) is an extension of the HA that aims to capture anharmonicity by introducing the volume dependence of frequencies due to temperature and pressure effects. In this work, we use the Phonopy package \cite{TOGO20151} version 2.29.0 to study thermal and mechanical properties of MOFs by studying the lattice dynamics under the QHA. A $11 \times 11 \times 11$ k-point grid, beyond the $\Gamma$ point, is consistently used for all phonon calculations,  which is essential to accurately capture heat transport properties.

\subsection{Overview of the investigated systems and potentials for benchmarking}
In this work, we benchmark the MACE-MP-MOF0 and MACE-MP-MOF0-v2 models against DFT, extended tight-binding DFT, MLPs and experiments. Among MLPs, we focus on a wide range of models from MACE-MP-0b (with D3 dispersion) to MTP and VASP MLP for MOFs \cite{Wieser2024}. For extended tight-binding DFT, we use the GFN1-xTB model \cite{D2CP00184E}. The GFN1-xTB model was built with a focus on the computation of molecular geometries, vibrational frequencies, and non-covalent interaction energies which are all important for this study of MOFs.

The MOFs chosen for benchmarking the phonons calculated with the MACE-MP-MOF0 models are widely studied and hence, enables thorough comparison of performance across the aforementioned models and experiments reported in literature. The investigated systems of MOF-5 (Zn), UiO-66 (Zr), MOF-74 (Zn) and MIL-53 (Al) are diverse in their nodes, linkers and topology (refer Table \ref{tab: lattice}). In this work, we study the large-pore configuration of the flexible framework, MIL-53 (Al). The number of reference configurations of these MOFs in the training sets of the two versions of MACE-MP-MOF0 models help analyze the transferability of the model and the performance dependence on the choice and size of training set. In addition to the primary benchmarking of phonon density of states and band structures for these four MOFs, benchmarking of the phonon-derived bulk modulus is done for several other well-known MOFs in literature which are not a part of the curated dataset for demonstrating the transferability of the MACE-MP-MOF0 model.
\begin{table*}[t]
    \centering
    \begin{tabular}{|c|c|c|}
    \hline
    MOF & MACE-MP-MOF0 & MACE-MP-MOF0-v2  \\
    \hline \hline
    MOF-5 (Zn2+) & 114 & 92 \\ \hline
    UiO-66 (Zr4+) & 68 & 58 \\ \hline
    MOF-74 (Zn2+) & 59 & 60 \\ \hline
    MIL-53 (Al3+) & 30 & 29 \\ \hline
    \end{tabular}
    \caption{The number of reference structures generated for the investigated MOFs for benchmarking the MACE-MP-MOF0 models in their respective training sets}
    \label{tab:train_mofs}
\end{table*}
\subsection{Benchmarking prediction of unit cell parameters of equilibrium structures}
\begin{figure*}[t]
    
    \begin{subfigure}[t]{0.50\textwidth}
        
        \includegraphics[width=\textwidth]{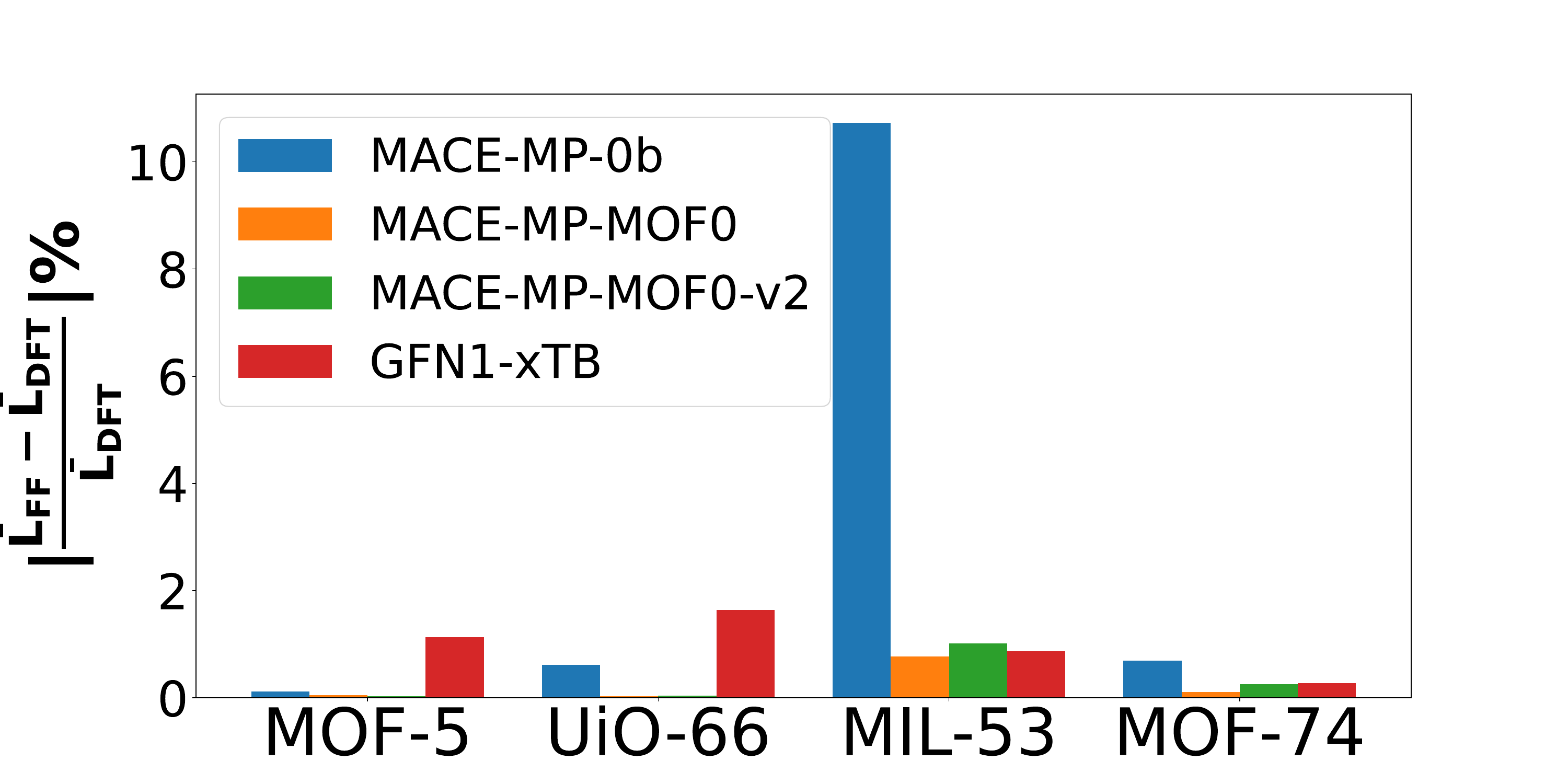}
        \caption*{\textbf{(a)}}
    \end{subfigure}
    \hfill
    \begin{subfigure}[t]{0.5\textwidth}
        
        \includegraphics[width=\textwidth]{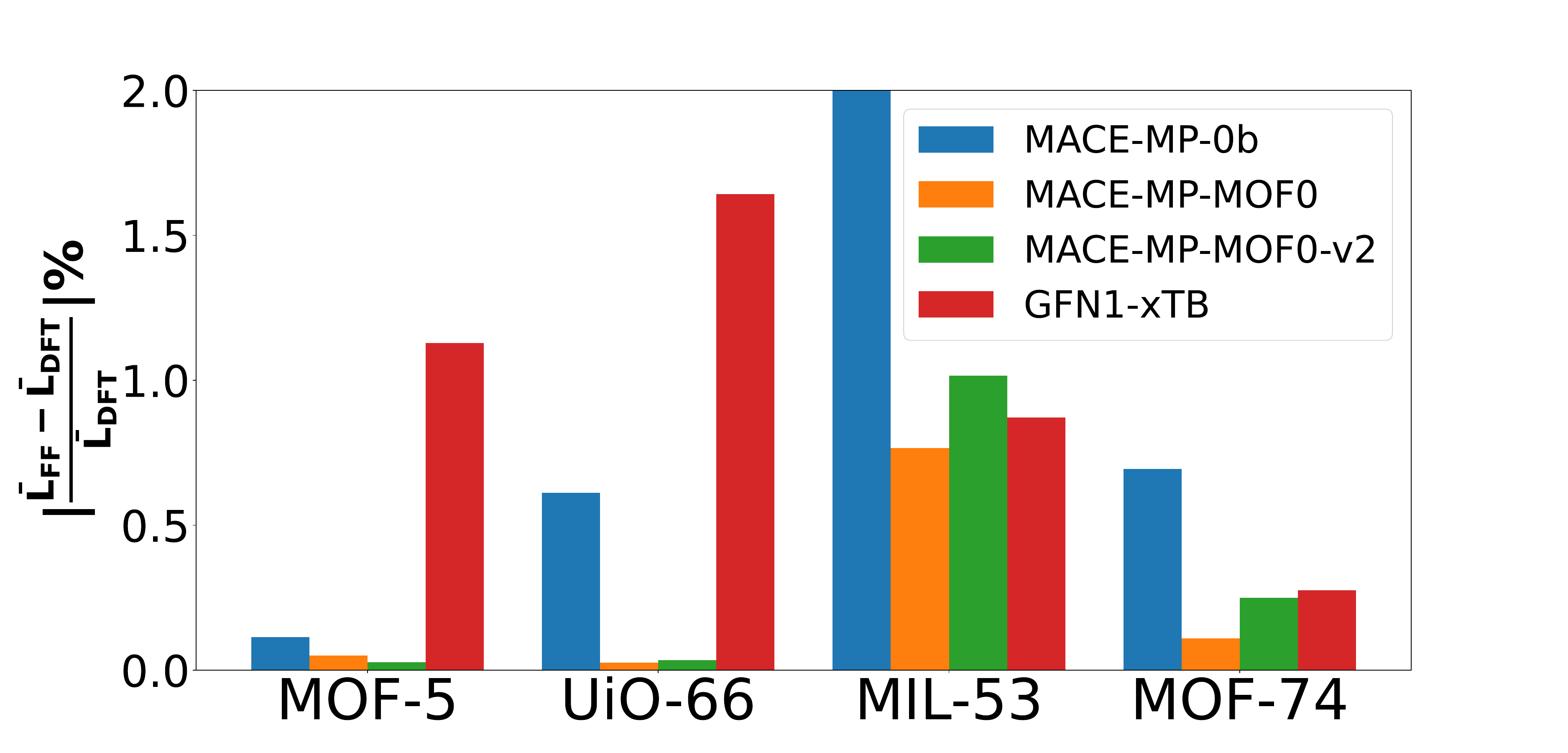}
        \caption*{\textbf{(b)}}
    \end{subfigure}
    
    \caption{Comparison of \% deviations in force field predictions of average equilibrium unit cell lengths $L_{FF}$ relative to DFT $L_{DFT}$ where FF stands for different models represented by bars of their respective colors for the investigated systems a) full range b) zoomed for better visualization of small deviations}
    \label{fig: lattice}
    
\end{figure*}

The first crucial step of benchmarking phonons is to accurately capture the equilibrium structure of the MOFs. Predictions of both MACE-MP-MOF0 and MACE-MP-MOF0-v2 show excellent agreement with DFT with a largest deviation of $1.02\%$ as shown in Figure \ref{fig: lattice}. The MACE-MP-0b deviations are much larger, with the largest deviation of 10\%. 

As shown in Table \ref{tab: lattice}, MACE-MP-MOF0 and MACE-MP-MOF0-v2 also predict the right space group for all MOFs while GFN1-xTB and MACE-MP-0b transform the unit cell to different space groups after a full cell relaxation. Even though GFN1-xTB produces deviations $\leq$ $2\%$ which are in good agreement with DFT, the minor distortions in unit cell lengths and angles shown in Table \ref{tab: lattice}, cause the identification of a different symmetry. As a result, obtaining phonons with GFN1-xTB for a highly symmetric structure like MOF-5 incurs a 17-fold increase in computational cost compared to the MACE-MP-MOF0 models. Symmetrizing the optimized unit cell by loosening the symmetry search tolerance does not eliminate the distortions and does not achieve the true symmetry with GFN1-xTB (see SI). Such symmetrizing procedures are not needed with the MACE-MP-MOF0 models,  which enables a full cell relaxation. Hence, in this work, the equilibrium structure is obtained by allowing the relaxation of only atomic positions with GFN1-xTB for phonon calculations.

\begin{table*}[t]
\centering
\renewcommand{\arraystretch}{1.2}
 \resizebox{\textwidth}{!}{
    \begin{tabular}{|c|c|c|c|c|c|}
    \hline
Method & MOF & Space Group & Unit Cell Lengths & Unit Cell Angles & No. of Symmetry \\
& &Symbol & a,b,c (\AA) & $\alpha$, $\beta$, $\gamma$ & Inequivalent Displacements\\
    \hline
    \hline
DFT & MOF-5& Fm-3m & 18.433, 18.433, 18.433 & 60.000, 60.000,60.000 & 19 \\
& UiO-66 & F-43m & 14.803, 14.803, 14.803 & 60.000, 60.000, 60.000 & 126 \\
& MIL-53 & Cc & 6.664, 17.263, 12.267 &  90.000, 90.000, 90.986 & 114 \\
& MOF-74 & R-3 & 6.665, 15.272, 15.272 & 62.088, 81.634, 98.366 & 162 \\
\hline
GFN1-xTB & MOF-5 & R-3m & 18.225, 18.225, 18.225 & 59.204, 59.204, 59.204 & 318 \\
&UiO-66 & F-43m & 14.560, 14.560, 14.560 & 59.996, 59.997, 59.995 & 684 \\
& MIL-53 & P1 & 6.772, 17.519, 12.219 & 90.095, 89.939, 89.643 & 456 \\
& MOF-74 & P1 & 6.693, 15.184, 15.230 & 62.284, 81.561, 98.497 & 324 \\
\hline
MACE-MP-0b & MOF-5 & Fm-3m & 18.412, 18.412, 18.412 & 60.000, 60.000, 60.000 &  19 \\
& UiO-66  & F-43m  & 14.713, 14.713, 14.713& 60.000, 60.000, 60.000 &  126 \\
& MIL-53 & C2/c  & 6.665, 19.447, 6.198 & 90.000, 90.000, 95.155 & 456 \\
&  MOF-74 & R-3 & 6.453, 15.249, 15.249 & 61.956, 81.891, 98.109 & 54 \\ 
\hline
MACE-MP-MOF0 & MOF-5 & Fm-3m &18.424, 18.424, 18.424 & 60.000, 60.000, 60.000 & 19 \\
 & UiO-66 & F-43m &14.799, 14.799, 14.799 & 60.000, 60.000, 60.000 & 34 \\
& MIL-53 & Cc & 6.669, 16.703, 13.099& 90.000, 90.000, 90.331& 228 \\
 & MOF-74 & R-3 & 6.692, 15.238, 15.238& 62.105, 81.582, 98.418 & 54 \\
 \hline
MACE-MP-MOF0-v2 & MOF-5 & Fm-3m & 18.428, 18.428, 18.428 & 60.000, 60.000, 60.000& 19 \\
& UiO-66& F-43m & 14.798, 14.798, 14.798 & 60.000, 60.000, 60.000 & 34 \\
& MIL-53 & Cc & 6.673, 17.795, 11.357 & 90.000, 90.000, 90.950 & 228 \\
& MOF-74 & R-3  & 6.654, 15.231, 15.231 & 62.083, 81.627, 98.373 & 54
\\ 
\hline\hline
    \end{tabular}
    }
    \caption{Comparison of optimized lattice parameters obtained from different models after a full cell relaxation of the investigated systems showing that the minor distortions up to the thousandths place in lattice parameters leads to significantly larger number of symmetry inequivalent displacements}
    \label{tab: lattice}
\end{table*}
\subsection{Benchmarking predictions of forces, energies and stress}
As phonon dispersion curves and density of states are obtained from force constants, an accurate description of interatomic forces in MOFs is necessary. Table \ref{tab:train_efs} shows the performance of both MACE-MP-MOF0 and MACE-MP-MOF0-v2 in obtaining the forces, energies and stress on their respective test sets. A linear fit with an $R^{2}$ score $\geq$ 0.999 is obtained for forces, energies and stress with MACE-MP-MOF0 and MACE-MP-MOF0-v2 relative to DFT for their respective training, test and validation sets (refer SI for figures).

The analysis of element-wise mean absolute errors (MAE) in energies relative to DFT for the MOFs in the QMOF database containing the 24 elements in the curated dataset reveal that the MACE-MP-0b predictions were 80\% larger than the MACE-MP-MOF0-v2 MAE predictions of 0.016 eV/atom (see SI). The MACE-MP-MOF0-v2 model shows significant improvement of 50\% in the energy MAE for MOFs with metal nodes like Zn, Al, Mg (which constitute a large set in the QMOF database) and in Hf and Zr-based MOFs (which are widely studied due to their importance in defect engineering of MOFs \cite{defect}). In this subset of the QMOF database, more than 75\% of MOFs show improved predictions, with the remaining 25\% primarily coming from the Cu-MOFs. The energy MAEs for integral elements of organic linkers like C, H and O common to all MOFs are also approximately lowered by 36\% with MACE-MP-MOF0-v2. Hence, overall we achieve a significant improvement in the MAE for the majority of the 24 elements in the curated MOFs, which cover 60\% of the QMOF database, by only including a small number of their reference configurations in the training dataset.

In addition to benchmarking the MACE-MP-MOF0 models for the curated dataset, we evaluate its performance on forces, as well as energies, on a set of 70 more diverse MOFs unseen by the model during training (see SI for structure details). As shown in Table \ref{tab:test_efs}, the MACE-MP-MOF0 models exhibit energy RMSDs relative to DFT that are five times lower than those of MACE-MP-0b, while also delivering 30\% more accurate forces and stresses for these out-of-sample MOFs.
We observe the deviations on out-of-sample MOFs to be 10 times the RMSDs of MACE-MP-MOF0 models on the curated dataset in Table \ref{tab:train_efs}, which demonstrates the significant improvement that can be achieved by adding a few reference configurations of MOFs to the training DFT data.
\begin{table*}[t]
    \centering
  
    \renewcommand{\arraystretch}{1.2}
    \begin{tabular}{|c|c|c|c|c|}
    \hline
    Model & $RMSD_{Energy}$ &$RMSD_{Energy/atom}$ & $RMSD_{Forces}$ & $RMSD_{Stress}$\\
    & (eV) & (meV/atom) & (eV/\AA) & (meV/\AA$^{3}$)\\
    \hline \hline
    MACE-MP-MOF0 & 0.132 & 1.4 & 0.014 & 0.2\\
    \hline
    MACE-MP-MOF0-v2 & 0.163& 3.6& 0.029& 0.3 \\
    \hline \hline
    \end{tabular}
    \caption{Root Mean Square Deviations (RMSDs) of forces, energies and stress predicted by MACE-MP-MOF0 and MACE-MP-MOF0-v2 relative to DFT on their respective test sets sampled from the curated dataset}
    \label{tab:train_efs}
\end{table*}
\begin{table*}[t]
    \centering
  
    \renewcommand{\arraystretch}{1.2}
    \begin{tabular}{|c|c|c|c|c|}
    \hline
    Model & $RMSD_{Energy}$ &$RMSD_{Energy/atom}$ & $RMSD_{Forces}$ & $RMSD_{Stress}$\\
    & (eV) & (meV/atom) & (eV/\AA) & (meV/\AA$^{3}$)\\
    \hline \hline
    MACE-MP-MOF0 & 0.773 & 12 & 0.146 & 1.1\\
    \hline
    MACE-MP-MOF0-v2 & 0.792& 10 & 0.171& 1.4 \\
    \hline
    MACE-MP-0b & 3.543 & 53 & 0.200 & 1.7 \\
    \hline \hline
    \end{tabular}
    \caption{Root Mean Square Deviations (RMSDs) of forces, energies and stress predicted by MACE-MP-MOF0, MACE-MP-MOF0-v2 and MACE-MP-0b relative to DFT on a diverse dataset consisting of 70 out-of-sample MOFs obtained from the QMOF database}
    \label{tab:test_efs}
\end{table*}

\subsection{Benchmarking vibrational properties}
\begin{figure*}[t]
    
    \begin{subfigure}[t]{0.50\textwidth}
        
        \includegraphics[width=\textwidth]{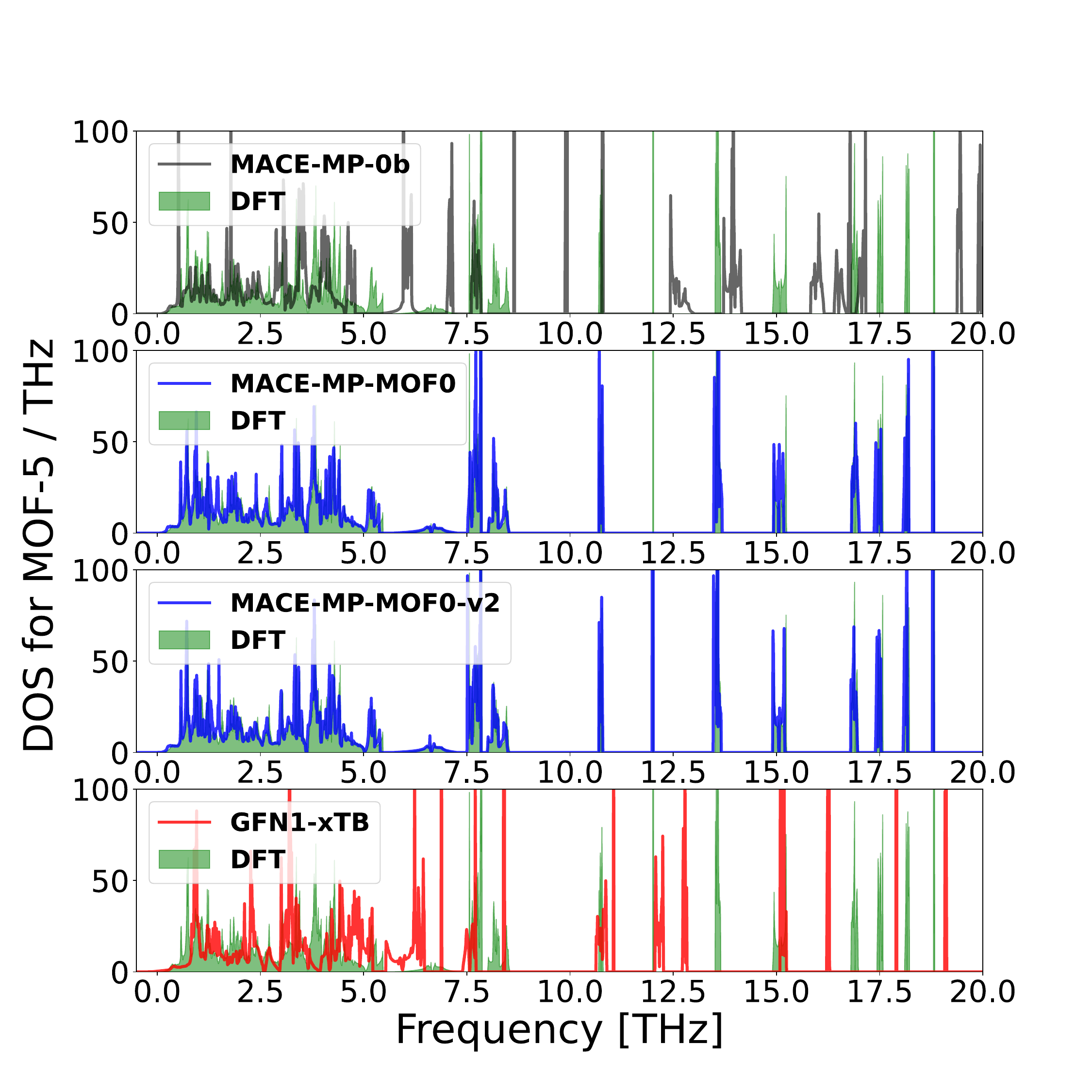}
        \caption*{\textbf{(a)}}
        \label{fig:doszoom1}
    \end{subfigure}
    \hfill
    \begin{subfigure}[t]{0.50\textwidth}
        
        \includegraphics[width=\textwidth]{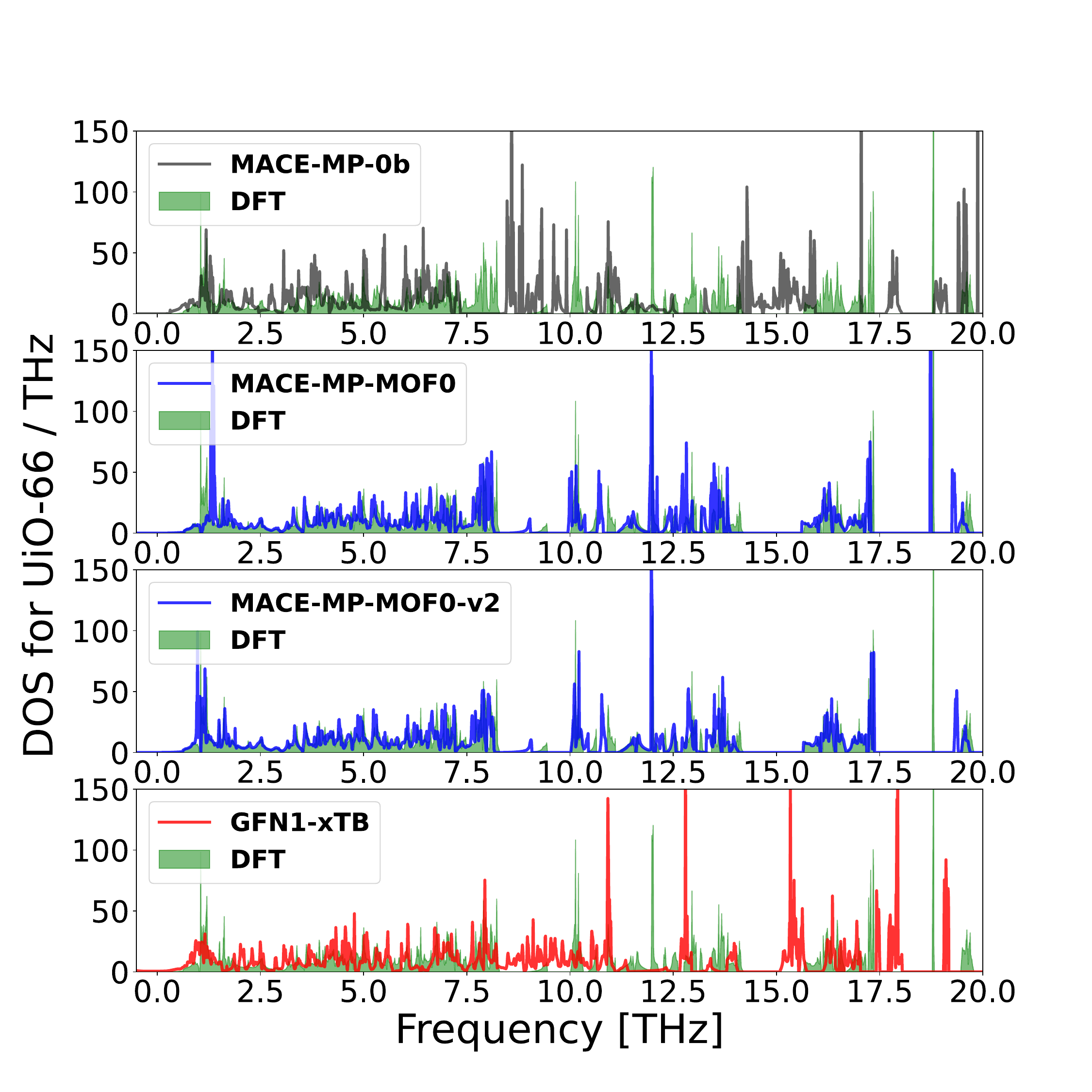}
        \caption*{\textbf{(b)}}
        \label{fig:doszoom2}
    \end{subfigure}
    \hfill
    \begin{subfigure}[t]{0.50\textwidth}
        
        \includegraphics[width=\textwidth]{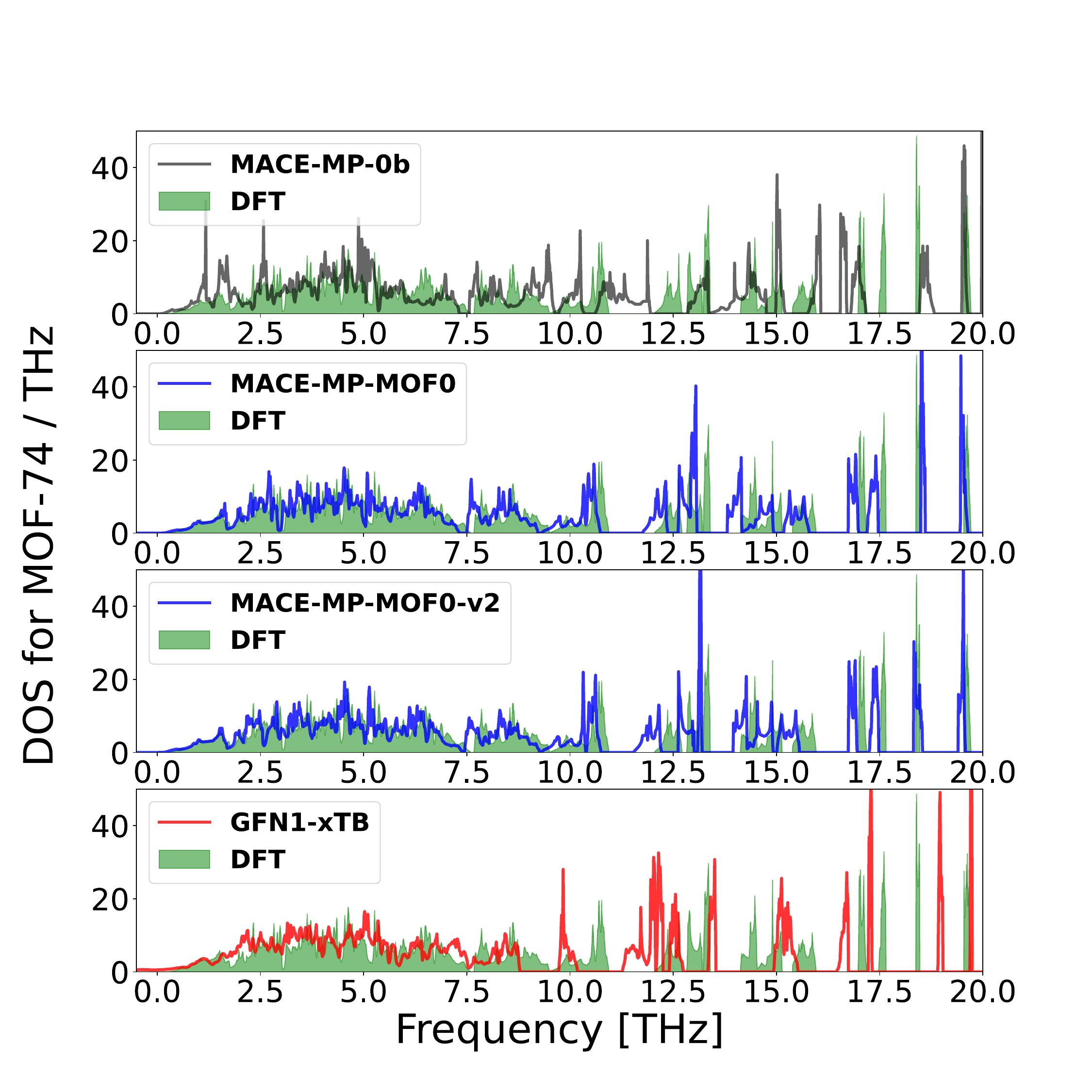}
        \caption*{\textbf{(c)}}
        \label{fig:doszoom3}
    \end{subfigure}
    \hfill
    \begin{subfigure}[t]{0.50\textwidth}
        
        \includegraphics[width=\textwidth]{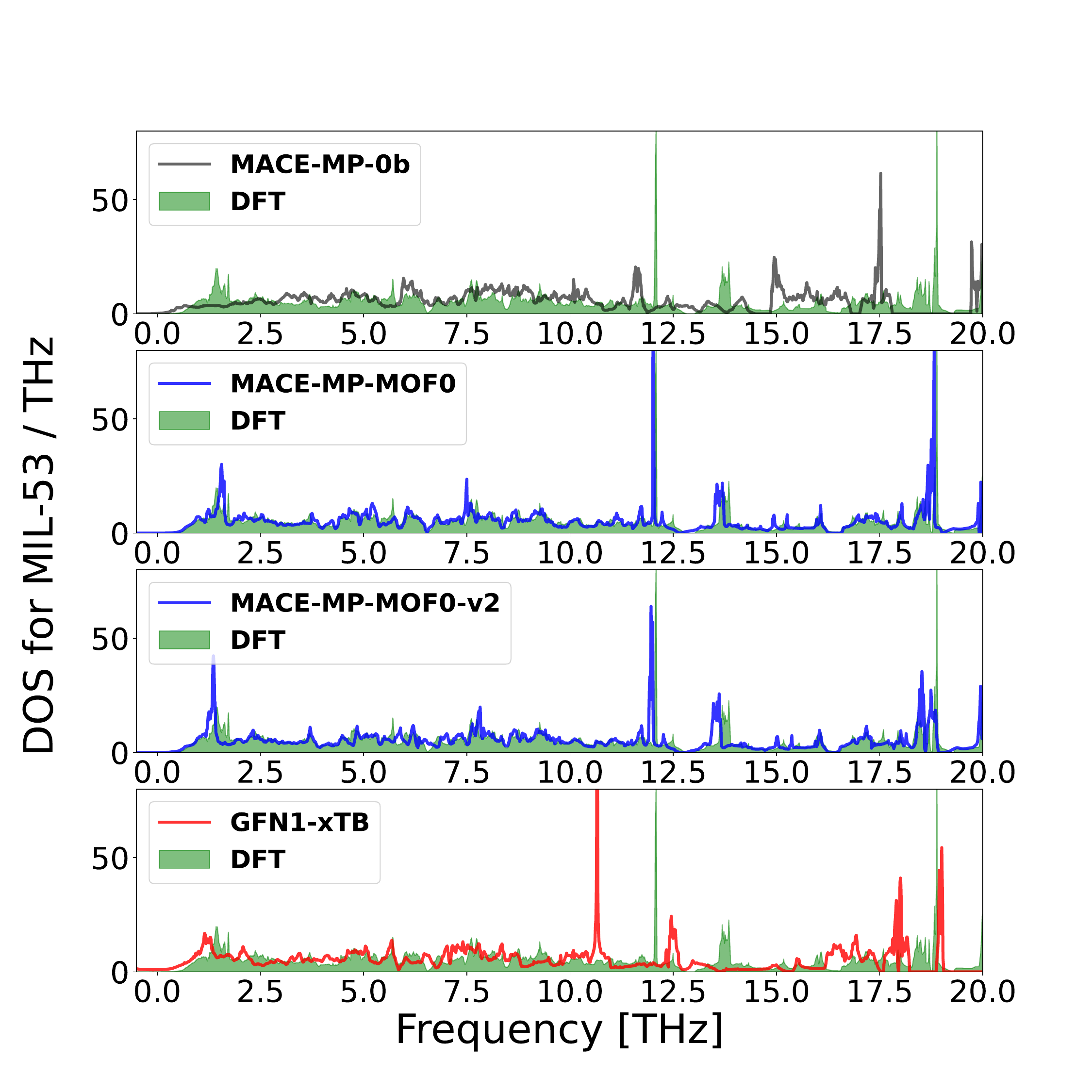}
        \caption*{\textbf{(d)}}
        \label{fig:doszoom4}
    \end{subfigure}
    \caption{Density of states (DOS) in the frequency range ($\leq$ 20 THz) for the investigated MOFs obtained with the linear tetrahedron method with a pitch of 0.01 THz predicted by the respective methods overlaid on DFT data. DFT DOS were calculated from force constants available in \cite{dft-mof-74,Wieser2024} on a $11\times11\times11$ mesh} 
    \label{fig: dos1}
\end{figure*}

After benchmarking forces, energies and stresses we  quantitatively analyze the derived phonon density of states and band structures. In Figure \ref{fig: dos1}, phonon density of states (DOS) data is shown up to 20THz (see SI for the full range of predicted frequencies). The MACE-MP-MOF0 and MACE-MP-MOF0-v2 models are in very good agreement with DFT DOS, even well above the low frequency range ($\leq$ 6THz). We were able to eliminate all spurious imaginary modes with the MACE-MP-MOF0 models to obtain physically meaningful frequencies. While GFN1-xTB and MACE-MP-0b capture certain aspects of the DOS curves accurately, they fail to accurately capture the vibrational properties of the most prototypical MOFs like MOF-5 and UiO-66. 

Table \ref{tab:phonon} shows the performance of MACE-MP-MOF0 and MACE-MP-MOF0-v2 in accurately obtaining the phonon frequencies for the full frequency range over the entire Brillouin Zone relative to DFT. The low errors over the full mesh ensure an accurate prediction of mechanical stability and heat transport properties as they depend heavily on off-$\Gamma$ phonons \cite{Ziman2001}. For direct comparison with the phonon frequencies reported at $\Gamma$-only with MTP and VASP MLP in \cite{Wieser2024}, we also report the $\Gamma$-only predictions with our models in Table \ref{tab:phonon}. A $\Gamma$-only comparison reveals that the MACE-MP-MOF0 models achieve a better or comparable performance for MOF-5, UiO-66 and MIL-53 than MTP and VASP MLP, while the errors are larger for MOF-74. It is important to emphasize that achieving this level of accuracy with such on-the-fly MLPs requires generating extensive DFT datasets tailored to specific MOFs. On the other hand, the MACE-MP-MOF0 training set had 10 to 100 times lower number of reference configurations for the respective MOFs, as indicated in Table \ref{tab:train_mofs}, than the training set for these on-the-fly MLPs \cite{Wieser2024}. The DFT data used here for benchmarking phonons, as well as the training data for the MTP and VASP MLP was generated in \cite{Wieser2024} with an extremely tight convergence criteria of $\text{ENCUT} = \text{900 eV}, \text{EDIFF} = 1 \times 10^{-8} \text{ eV with a max force component} \leq 1 \times 10^{-3} \text{ ev/\AA }$, which are necessary to obtain accurate phonons rendering the DFT calculations very expensive. Whereas, the training DFT data generated in this work for MACE-MP-MOF0 models used a significantly cheaper convergence criteria of $\text{ENCUT} = \text{520 eV, EDIFF} = 1 \times 10^{-5} \text{ eV and EDIFFG} = \text{-0.02 eV/\AA }$ (refer to the Methods section for details). This demonstrates that the fine-tuned transferable MLP, MACE-MP-MOF0, can achieve the same high-quality ab initio description as other accurate machine-learned force field potentials for MOFs \cite{Wieser2024,Vandenhaute2023,mlff}, but with significantly reduced model training and DFT data generation costs.

\begin{figure*}[t]
    
    \begin{subfigure}[t]{0.50\textwidth}
        
        \includegraphics[width=\textwidth]{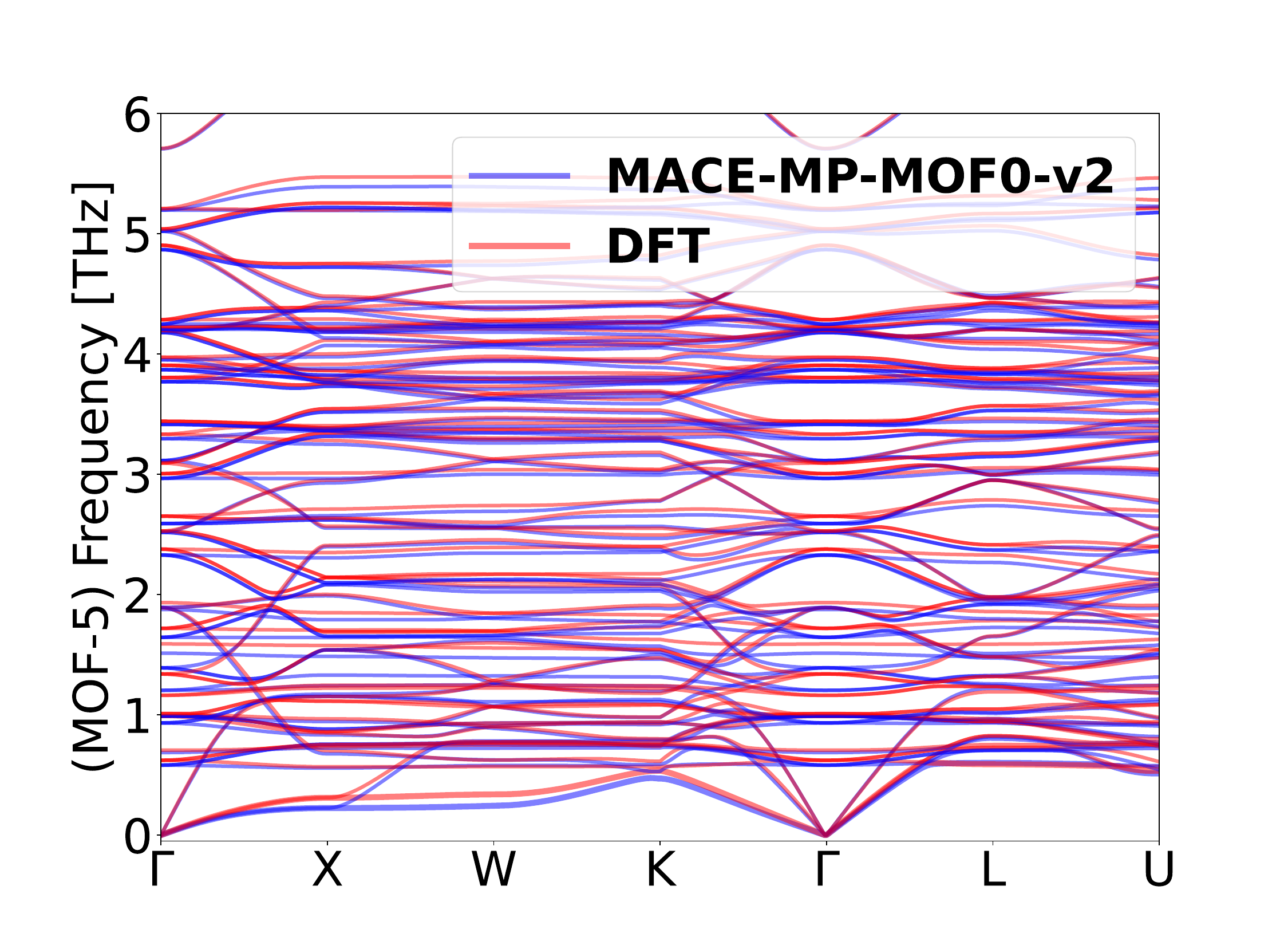}
        \caption*{\textbf{(a)}}
        \label{fig:bszoom1}
    \end{subfigure}
    \hfill
    \begin{subfigure}[t]{0.50\textwidth}
        
        \includegraphics[width=\textwidth]{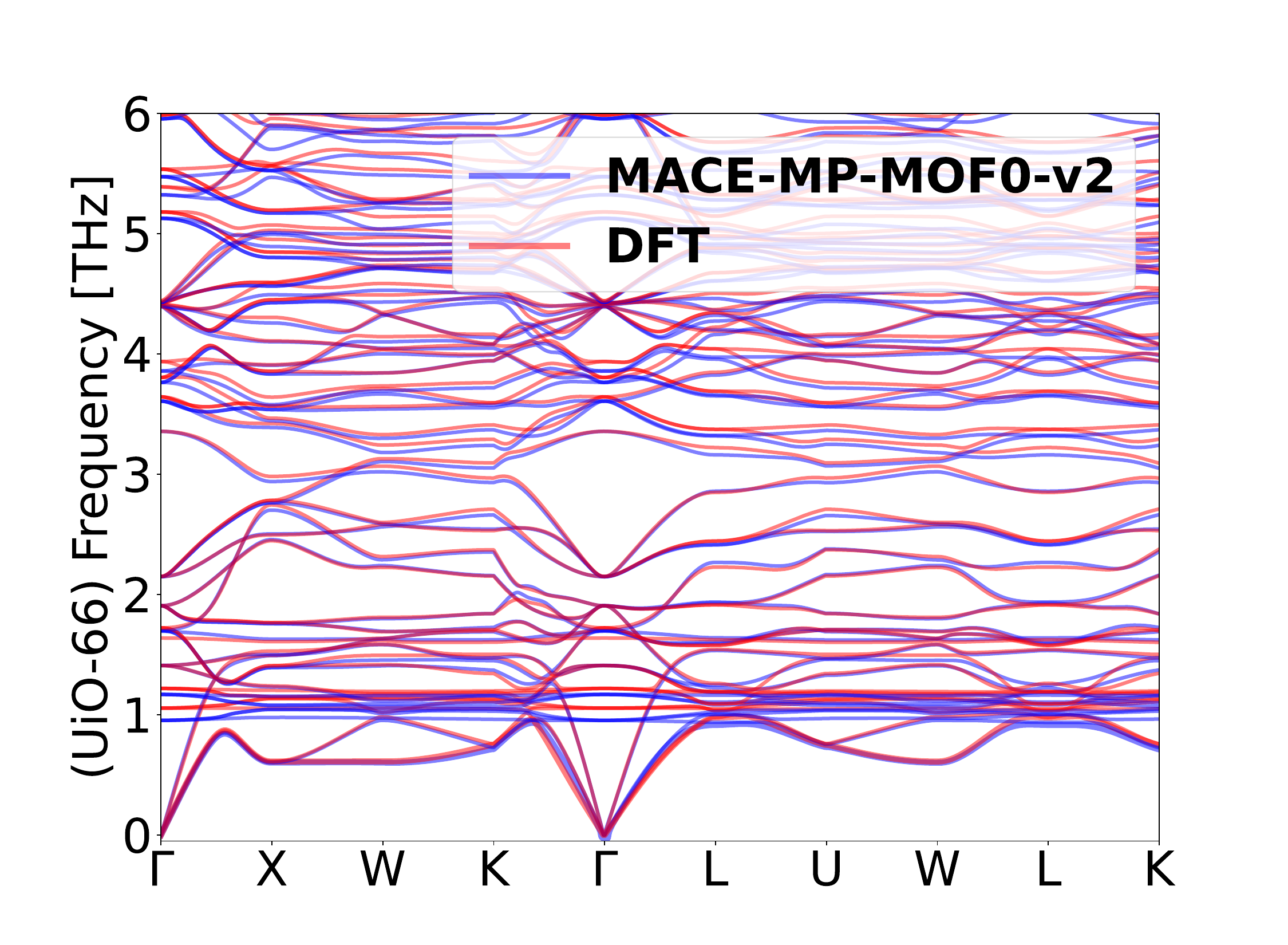}
        \caption*{\textbf{(b)}}
        \label{fig:bszoom2}
    \end{subfigure}
    \hfill
    \begin{subfigure}[t]{0.50\textwidth}
        
        \includegraphics[width=\textwidth]{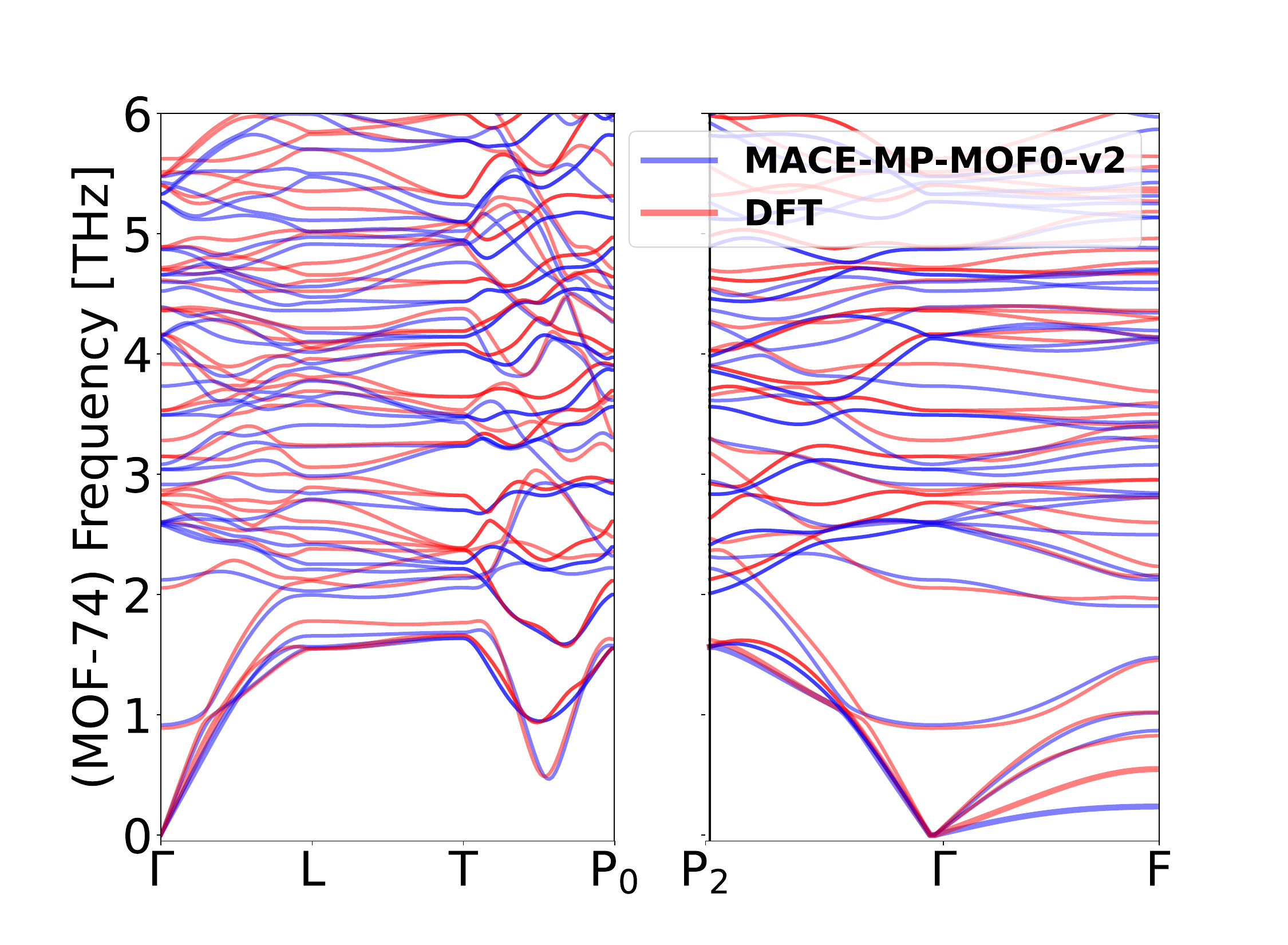}
        \caption*{\textbf{(c)}}
        \label{fig:bszoom3}
    \end{subfigure}
    \hfill
    \begin{subfigure}[t]{0.50\textwidth}
        
        \includegraphics[width=\textwidth]{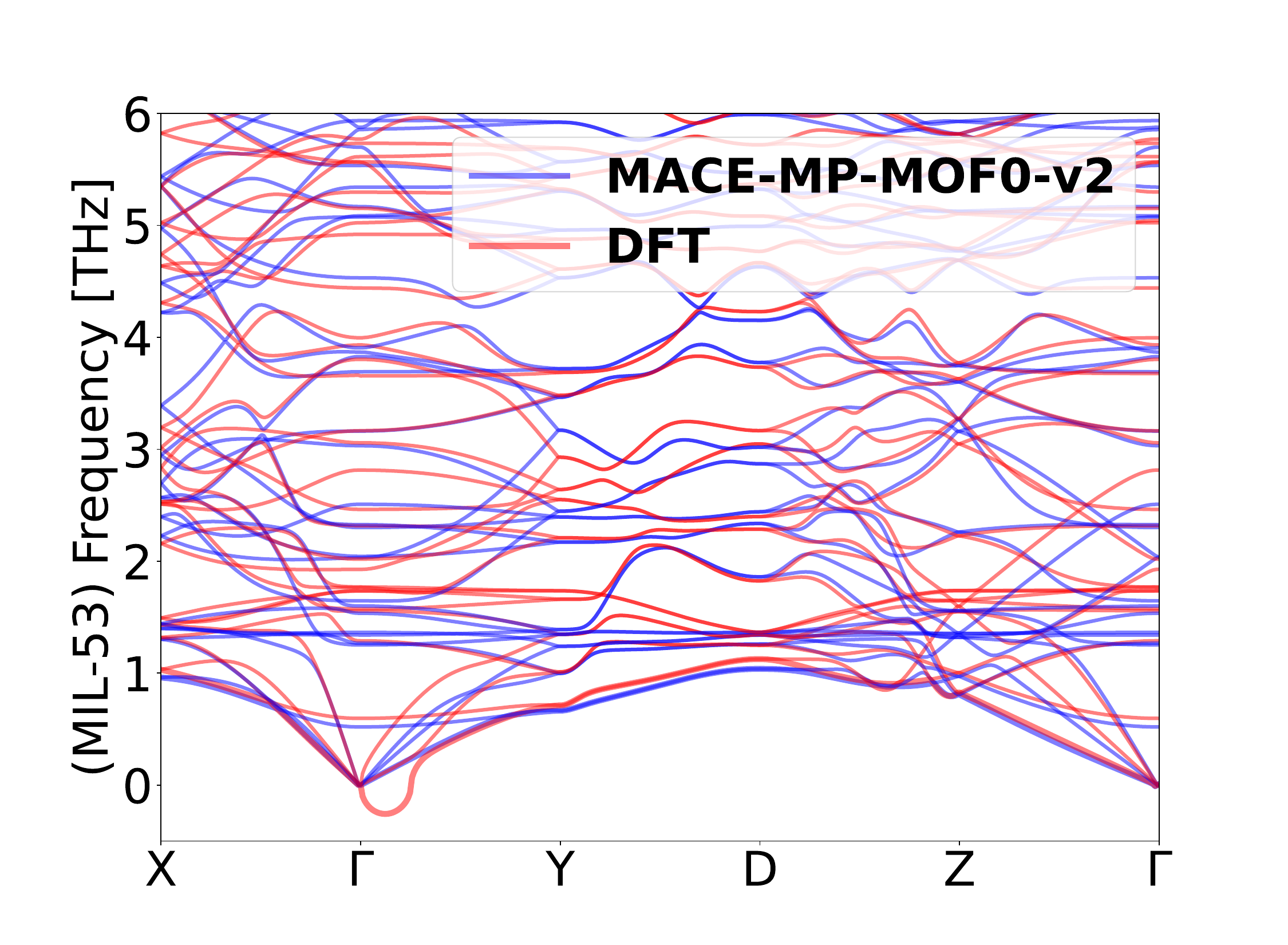}
        \caption*{\textbf{(d)}}
        \label{fig:bszoom4}
    \end{subfigure}
    \caption{Low frequency band structures ($\leq$ 6THz) predicted by MACE-MP-MOF0-v2 overlaid on DFT band structure for the investigated MOFs. DFT band structures were calculated from force constants available in \cite{dft-mof-74,Wieser2024}}
    \label{fig: bs}
\end{figure*}

The similar errors obtained with MACE-MP-MOF0 and MACE-MP-MOF0-v2 in Table \ref{tab:phonon} show that the different reference configurations of the investigated systems in their respective training sets does not significantly affect the performance of the models. Since MACE-MP-MOF0 and MACE-MP-MOF0-v2 models exhibit similar frequency errors, we use the slightly better performing MACE-MP-MOF0-v2 model for these MOFs to show the accurate overlap between the band structures predicted with DFT and MACE-MP-MOF0-v2 in Figure \ref{fig: bs}. Due to the large number of bands in MOFs, the band structures in figure \ref{fig: bs} are restricted to the low frequency region.

\begin{table*}[t]
    \centering
    \renewcommand{\arraystretch}{1.2}
    \begin{tabular}{|c|c|c|c|c|}
    \hline
    MOF & $Full\_Mesh_{\text{MACE-MP-MOF0-v2}}$ & $Full\_Mesh_{\text{MACE-MP-MOF0}}$ & $\Gamma-only_{\text{MTP}}$ & $\Gamma-only_{\text{VASP MLP}}$ \\
 & (THz) & (THz) & (THz) & (THz) \\
 \hline \hline
MOF-5 & 0.033 (0.085) & 0.039 (0.083) & 0.099 & 0.27 \\
\hline
UiO-66 & 0.082 (0.186) & 0.122 (0.198) & 0.111	& 0.312 \\
\hline
MOF-74	& 0.173 (0.427)	& 0.243 (0.436)	& 0.093	& 0.126 \\
\hline
MIL-53 & 0.138 (0.257)	& 0.126 (0.239) &	0.156	& 0.264 \\
\hline \hline
    \end{tabular}
    \caption{Comparison of Root Mean Square Deviations (RMSDs) of the phonon frequencies in the full range for the investigated MOFs with MTP and VASP MLP data as reported in \cite{Wieser2024}. $\Gamma$-only errors with MACE-MP-MOF0 models are provided in brackets.}
    \label{tab:phonon}
\end{table*}
\subsection{Benchmarking of mechanical properties}
\begin{figure*}[t]
    
    \begin{subfigure}[t]{0.50\textwidth}
        
        \includegraphics[width=\textwidth]{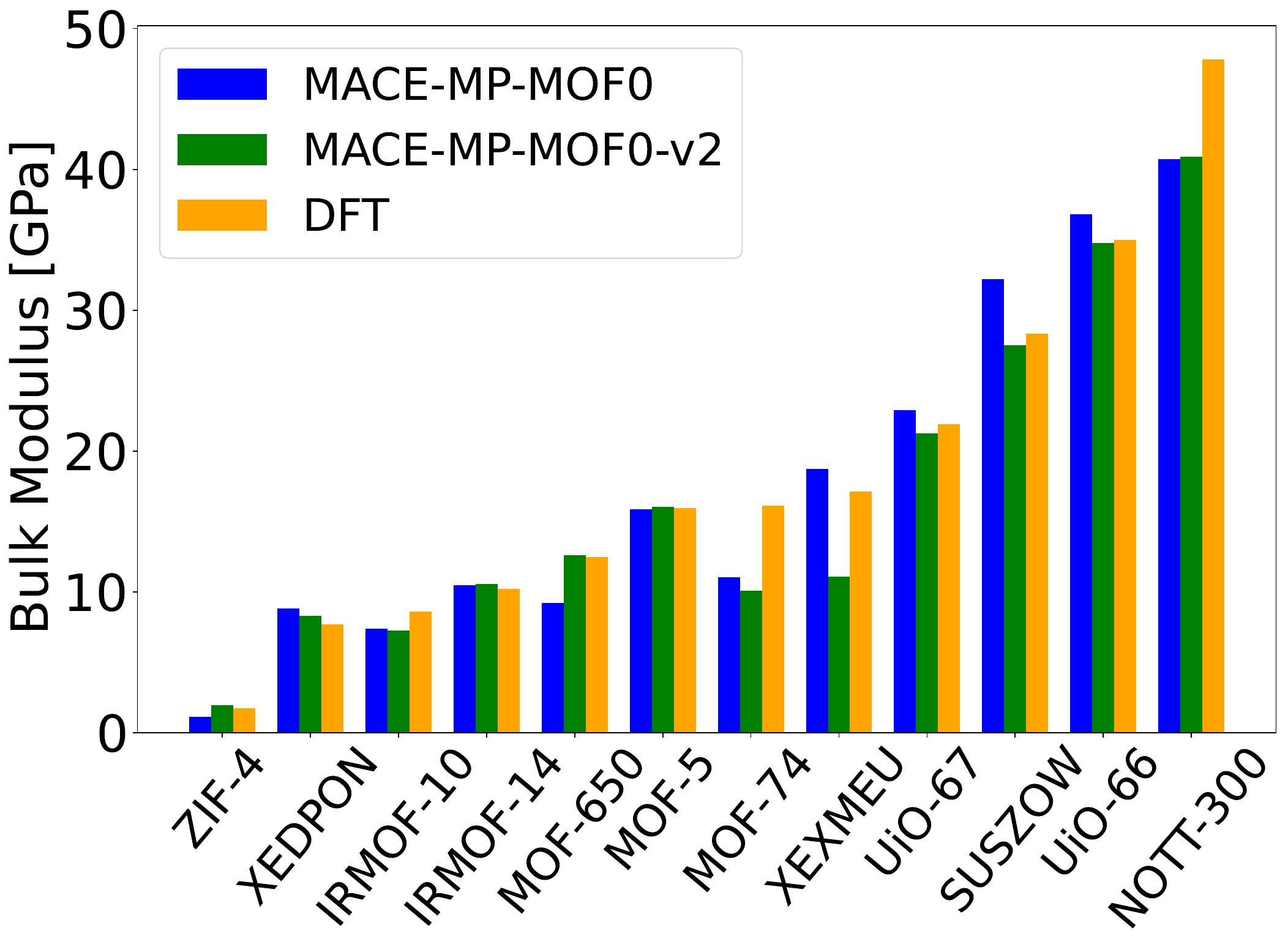}
        \caption*{\textbf{(a)}}
        \label{subfig:dft}
    \end{subfigure}
    \hfill
    \begin{subfigure}[t]{0.45\textwidth}
        
        \includegraphics[width=\textwidth]{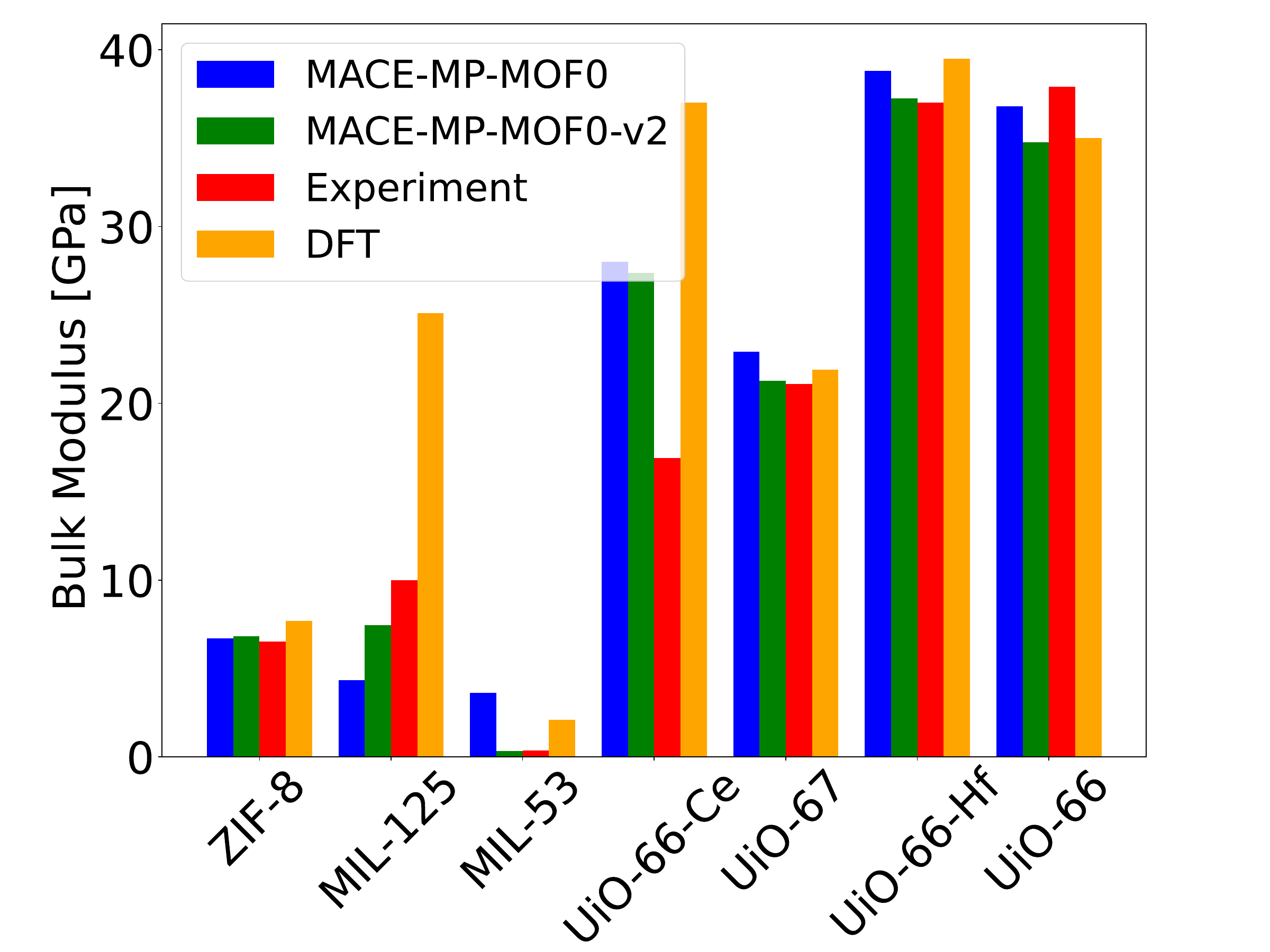}
        \caption*{\textbf{(b)}}
        \label{subfig:exp}
    \end{subfigure}
    \caption{Bulk modulus predictions of MACE-MP-MOF0 models as compared to a) only DFT and b) DFT and Experiments where ZIF-4, ZIF-8, XEDPON, XEXMEU, SUSZOW \cite{opendac}, UiO-66-Hf/Ce are diverse out-of-curated dataset MOFS. DFT data is self-generated or reported in \cite{Shi2022, irmof-14,C6CP05106E,hoffman53} and experiment reported in \cite{zif8,mil-125,mil-53,uio66-ce,uio67}}
    \label{fig: bulk}
\end{figure*}

The bulk modulus serves as a sensitive and computationally efficient metric for assessing phonon accuracy, as both phonon frequencies and bulk modulus arise from second derivatives of energy- phonons with respect to atomic displacements and bulk modulus with respect to volume. Since both properties share similar force accuracy requirements, an accurate bulk modulus indicates that the elastic response and interatomic potential curvature are well captured, suggesting reasonable phonon predictions. Hence, in this section, we analyze mechanical properties, such as the bulk modulus, for additional widely studied MOFs beyond the four systems investigated, comparing them to experimental and DFT data reported in the literature. The bulk modulus is obtained from fitting the Birch-Murnaghan equation of state to energy-volume data obtained by straining the cell $\pm 2\%$.
From Figure \ref{fig: bulk}, we observe that MACE-MP-MOF0 and MACE-MP-MOF0-v2 are able to qualitatively capture bulk moduli trends, compared to DFT and experimental data, as well as quantitatively reproduce the values with minor deviations. 

The deviations in bulk modulus with MACE-MP-MOF0 versions relative to DFT are the largest for MIL-125 in Figure \ref{fig: bulk} b), which notably, is the only Ti-based MOF in the curated dataset. We hypothesize that the deviation is due to insufficient training data for Ti-based MOFs. Serendipitously, the MACE-MP-MOF0 and MACE-MP-MOF0-v2 predictions of bulk moduli actually perform better than the explicit DFT data, relative to experiments. The DFT data includes VdW corrections, which tends to overestimate the long range interactions in MOFs and therefore produces higher bulk modulus than experiments \cite{longrange}. As the MACE-MP-MOF0 model was trained on DFT data with vdW corrections included, these long-range interactions are therefore slightly underestimated in the model relative to DFT, which moderates the bulk moduli predictions. Hence, the small RMSDs obtained for phonon frequencies with MACE-MP-MOF0-v2 and MACE-MP-MOF0 translate to low deviations in phonon-derived properties such as the bulk modulus.

The agreement between experiments, DFT and MACE-MP-MOF0 models in obtaining the bulk modulus of several diverse out-of-sample, well-known MOFs like ZIF-4, ZIF-8, UiO-66-Ce, UiO-66-Hf and promising MOFs for direct air capture like XEDPON, SUSZOW and XEXMEU \cite{opendac}, thus demonstrates the transferability and accuracy of the model in obtaining bulk modulus of MOFs unseen by the model during training.
\subsection{Negative Thermal Expansion in MOFs}
Finally, we evaluated our model on negative thermal expansion (NTE). Experimental data on NTE for MOFs is limited, making it challenging to benchmark against a wide range of models and structure. In addition, obtaining NTE with DFT is computationally expensive due to the quasi-harmonic phonon calculations required for MOFs. To address this challenge, in this work, we focused our NTE analysis on MOF-5 and UiO-66 as they are widely studied in experiments and their high symmetry and isotropic nature reduces the cost of such expensive computations. 

Comparing DFT and MACE-MP-MOF0-v2 predictions to experimental data is challenging, as factors such as pressure, cooling or heating rates, as well as defect concentration can vary significantly in experiments, influencing the recorded NTE \cite{C5CP01307K,dft-66}. For example, in \cite{dft-66}, the authors observed that, while the equilibrium UiO-66 intrinsically shows a coefficient of thermal expansion (CTE) of $\text{ -35 } \times 10^{-6} \text{ K}^{-1}$, a wide range of CTE from $ \text{+45 to -80 } \times 10^{-6} \text{ K}^{-1}$ is recorded depending on the rate of thermal treatment in UiO-66, suggesting a re-evaluation of previous experimental reports of NTE in MOFs. 

Based on the available data, we find that MACE-MP-MOF0-v2 predictions are in very good agreement with DFT and other computational models and qualitatively capture experimental NTE trends, such as the larger NTE observed for UiO-66 compared to MOF-5 (see Table \ref{tab:nte_mofs}). 

\begin{table*}[t]
   \centering
    \renewcommand{\arraystretch}{1.2}
    \begin{tabular}{|c|c|c|c|c|}
    \hline
    MOF & MACE-MP-MOF0-v2 & DFT& Simulation & Experiment \\
     & [$10^{-6} K^{-1}$] & [$10^{-6} K^{-1}$] & [$10^{-6} K^{-1}$] & [$10^{-6} K^{-1}$]\\
    \hline \hline
    MOF-5 (300K) & -6.65 & $-3.5^{(1)}$ \cite{nte-dft} & -8 \cite{nte-md} & -13.1 \cite{nte-exp} \\
    \hline
    UiO-66 (350K) & -9.35 &$-4^{(1)}$ \cite{C6CP05106E}&$-6^{(1)}$ \cite{C6CP05106E} & -5.3 to -35 \\
    & & & &((323K-523K),Rate of cooling) \cite{dft-66}
    \\
    \hline\hline
    \end{tabular}
    \caption{Comparison of CTE predictions by different methods at different specified temperature and experimental conditions,(1) indicate values read from graph}
    \label{tab:nte_mofs}
\end{table*}

\section{Discussion}
\begin{figure*}[t]
    \centering
    \includegraphics[width=0.5\linewidth]{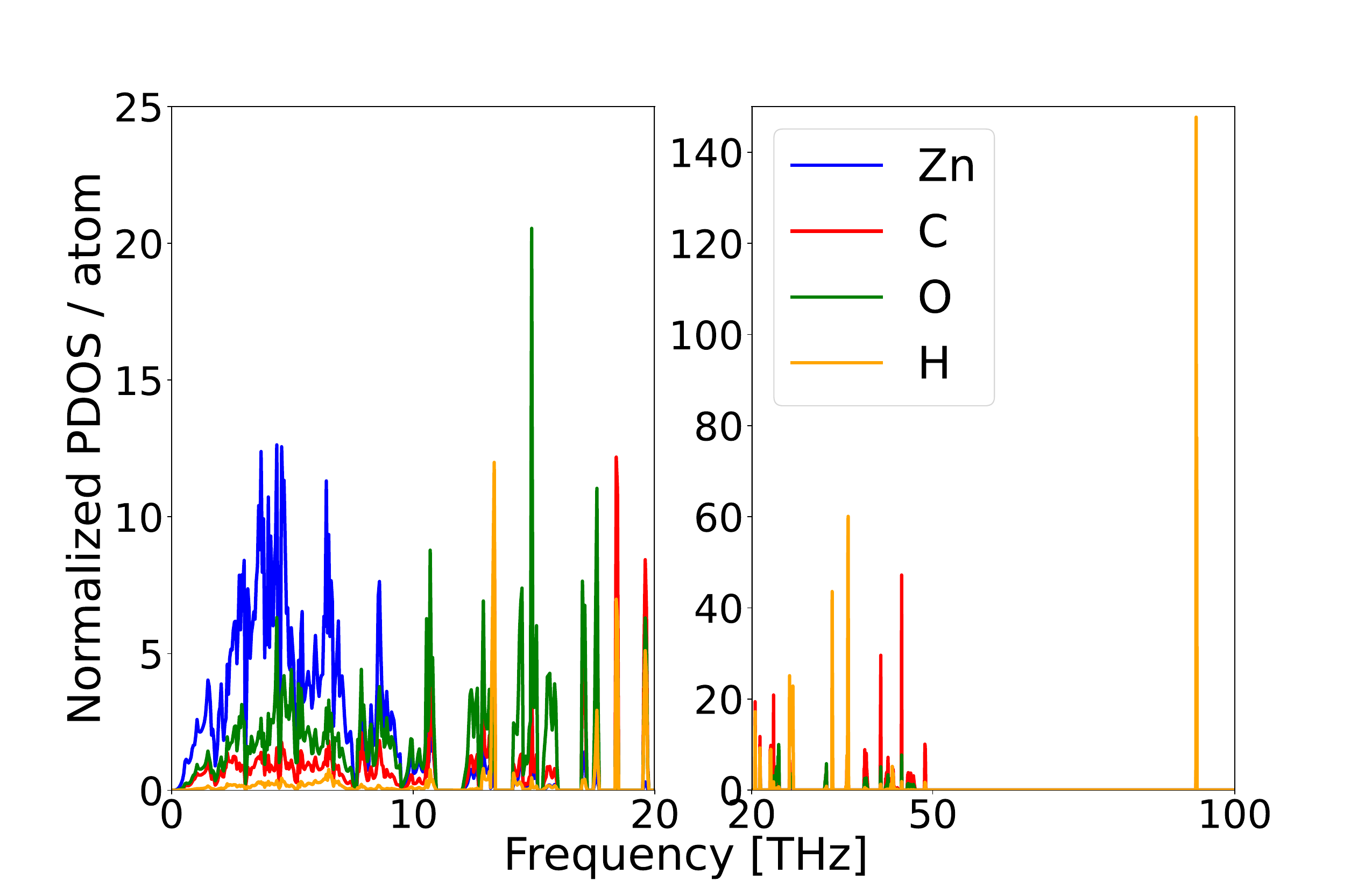}
    \caption{Normalized Partial Density of States (PDOS) for MOF-74 showing the contributions of the composing elements to the phonon spectrum}
    \label{fig:pdos}
\end{figure*}

Benchmarking the vibrational properties of the investigated MOFs highlights the limitations of MACE-MP-0b and GFN-1xTB, while demonstrating the excellent agreement of the MACE-MP-MOF0 models with DFT in accurately capturing the full phonon vibrational spectra. We further rationalize the performance of these models by analyzing their ability in accurately capturing the atomic positions and interactions in the equilibrium structure in Table \ref{tab:pos} for MOF-74. An accurate capture of atomic positions in the MOF provides a better description of forces, which ultimately correlates to the partial phonon density of states (PDOS) obtained with DFT in Figure \ref{fig:pdos}. 
  
Figure \ref{fig: dos1} for MOF-74 shows that the GFN1-xTB method is in good agreement with DFT  for frequencies lower than 6 THz,  above which the errors significantly increase. The overlap is  better in the lower frequency range because this region is dominated by the heavy Zn atom vibrations which have low RMSDs in Table \ref{tab:pos}. The GFN1-xTB and MACE-MP-0b errors in the atomic positions for C and O, dominating the vibrations beyond 6THz, are ten to hundred times larger than the MACE-MP-MOF0 errors. The greater flexibility and higher rotational degrees of freedom of these atoms in the organic linkers are more difficult to capture than the surrounding metal nodes \cite{D2CP00184E}.Therefore, the GFN1-xTB method accurately captures the vibrations of the metal nodes, but it inadequately represents the interactions between atoms in the organic linkers for phonon calculations. As the MACE-MP-MOF0 models consistently achieves ten to thousand times lower errors for all the elements present in the MOFs than GFN1-xTB and MACE-MP-0b, we obtain the high agreement between DFT and MACE-MP-MOF0 model predictions throughout the phonon spectra (see SI for the analysis of other MOFs). MACE-MP-MOF0 demonstrates a significant improvement in capturing the covalent interactions in MOFs, addressing the limitations of MACE-MP-0b and GFN1-xTB, which primarily capture non-covalent interactions. This enhancement positions MACE-MP-MOF0 as a promising tool for systems beyond MOFs, where covalent interactions dominate, such as in covalent organic frameworks.\cite{cof}.
\begin{table*}[t]
   \centering
    \renewcommand{\arraystretch}{1.2}
    \begin{tabular}{|c|c|c|c|c|}
    \hline 
    Method & $RMSD_{Zn}$ & $RMSD_{C}$ & $RMSD_{H}$ & $RMSD_{O}$ \\
    \hline \hline
    GFN1-xTB & 0.026 & 0.11 & 0.010 & 0.125\\
    MACE-MP-0b & 0.009 & 0.005 &0.005& 0.013 \\
    MACE-MP-MOF0 & 0.002 & 0.001& 0.001 & 0.002\\
    MACE-MP-MOF0-v2 & 0.003 & 0.002& 0.001 & 0.004 \\
    \hline \hline
    \end{tabular}
    \caption{Root Mean Square Deviations (RMSD)s of optimized fractional atomic positions with the respective methods relative to DFT }
    \label{tab:pos}
\end{table*}

In summary, in this work, we present a fine-tuned MLP for MOFs, MACE-MP-MOF0, that can be used to produce ab initio quality phonons in a high-throughput way. The model was trained on a representative dataset of 127 MOFs selected by efficiently sampling the phase space based on MACE descriptors which significantly reduced the computational efforts in generating DFT reference data. The MACE-MP-MOF0 model predicts phonon-derived bulk modulus in excellent agreement with experiments and DFT on MOFs unseen by the model, demonstrating its transferability.

The MACE-MP-MOF0 model presented here is also 50\% faster than the MACE-MP-0b foundation model (with dispersion corrections included) and 10 times more accurate in obtaining optimized geometries, forces, energies and stresses which are crucial for obtaining accurate vibrational properties. Additionally, MACE-MP-MOF0 model is 90\% faster than GFN1-xTB for large MOFs (500 atoms per unit cell) (see SI) The computational efficiency and accuracy of the model make it an excellent candidate for screening phonon-derived properties, such as bulk modulus, in MOFs, providing a quantitative description that aligns closely with experimental and DFT results. The model can calculate phonons within the quasi-harmonic approximation to obtain ab initio-level thermal expansions, which would otherwise be computationally expensive to obtain using DFT for MOFs.

\begin{figure*}
    \centering
    \includegraphics[width=0.5\linewidth]{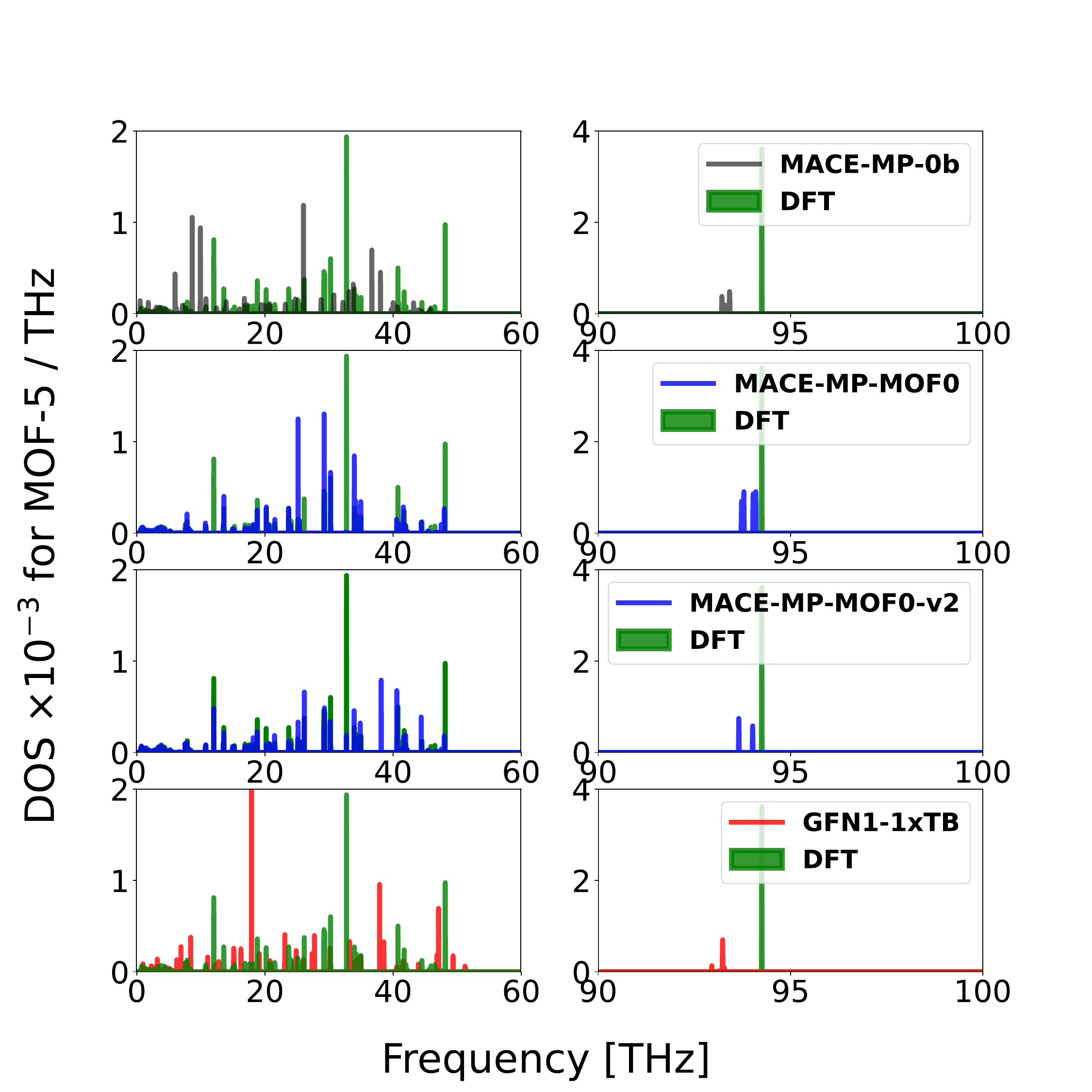}
    \caption{DOS for MOF-5 in the full frequency range}
    \label{fig:full}
\end{figure*}

Through the preliminary analysis for thermal expansion of MOF-5 and UiO-66 with MACE-MP-MOF0 and DFT, we conclude that the quasi-harmonic approximation does not provide a sufficient description of anharmonic effects in MOFs and is an area with scope for improvement in the computational study of vibrational properties of MOFs. While MACE-MP-MOF0 accurately predicts phonons in the majority of the frequency range (0 to 60 THz) which dominate most of the physical properties of interest, nevertheless, further improvement is possible by capturing the high frequency range (90 - 100 THz, which correspond to the vibration of H atoms in the linkers) as shown in Figure \ref{fig:full}. We would also like to highlight that while the current model covers 60\% of MOFs in the QMOF database that span the same chemical space of metal nodes as the 127 MOFs, these sampled MOFs are majority closed-shell metal ions, which avoids electronic spin-degrees of freedom. In MOFs with magnetic elements, having training data of atomic configurations with different spin-states in the same MOF can lead to poor training of the MLP. HKUST-1, a well-known Cu-based MOF, is the only MOF in the curated dataset with a magnetic element which contributes to the higher element-wise MAE obtained with MACE-MP-MOF0 for Cu (refer SI).  In addition to the model presented here being ready-to-use for a wide range of metallic nodes, organic linkers, and topologies, it is also significantly easier to re-parameterize for new species by simply including a few reference configurations of the MOF into the DFT training set, compared to traditional force fields.

In conclusion, the presented model provides a platform for performing high-throughput calculations in an efficient manner to guide the design and synthesis of MOFs for complex dynamical properties. The reported data motivates further development of MLPs that are easy-to-train and transferable to replace extremely expensive ab initio methods for the analysis of lattice dynamics in MOFs.

\section{Methods}
\subsection{DFT Calculations}
The DFT computations for generating the dataset were performed using VASP \cite{vasp,vasp1,vasp2,vasp3} and the atomate2 software\cite{atomate2}. Since MACE-MP-0b was trained on the MPtrj dataset \cite{mptrj}, we ensured consistency by using the same parameters and pseudo potentials to produce our DFT training dataset. The Perdew-Burke-Ernzerhof (PBE) form of the generalized gradient approximation (GGA) exchange-correlation functional \cite{pbegga} was used along with the Projector Augmented Wave (PAW) method (version 54) \cite{paw}. A D3(BJ) van der Waals correction \cite{bj,d3} was included for all MOFs, which eliminated some spurious phonon instabilities. In \cite{func}, it has been shown that different functionals perform similarly in structure prediction when applied to chemically diverse MOFs for screening purposes. Hence, the chosen PBE-D3 (BJ) functional offers computational consistency and convenience for the diverse curated dataset. Adding the dispersion correction to the training DFT data instead of adding dispersion on the MACE-MP-0b model increased the speed of the model by 1.5 times (refer SI).

Static DFT data was produced with an energy convergence of $10^{-5}$ eV. The cutoff energy for the plane-wave basis set was set to 520 eV. Geometry optimization was performed on deformed configurations of the curated MOFs to produce trajectories with DFT with a force convergence of $ \text{EDIFFG} = -0.02 $ ev/\AA. DFT data for Equation of State (EOS) was generated by obtaining the energies of 6 isotropic deformations with a linear deformation ranging from - 10\% to 10\% using the Birch-Murnaghan EOS. 

\subsection{Improving accuracy of MLP via Fine-tuning}
Fine tuning was carried out on our curated dataset using the multi-head approach as implemented in the MACE code starting with version 0.3.7. We used MACE-MP-0b as one head and our PBE+D3(BJ) \cite{pbegga,paw,d3,bj} calculations as the other head. The ratio of weights for the loss functions for energy, forces and stresses was 1:10:100 and we tuned the MLP training for 2500 epochs. It is important to note that during the training process, MACE employs a FPS approach to select configurations from the original MPtrj training set \cite{mptrj} and adds them to our data.  Table \ref{tab:train_efs} shows the final convergence results for the fine tuning for the two models trained, MACE-MP-MOF0 and MACE-MP-MOF0-v2.  Due to the relatively large number of atoms in our training set per MOF, training was carried out on 16 A100@40GiB RAM NVIDIA GPUs. The hyperparameters for fine tuning the model, are inherited from the original MACE-MP-0b model (L=1, $r\_max=6.0$, hidden irreps 128x0e+128x1o,  2 layers, each with correlation order: 3 (body order: 4) and spherical harmonics up to: l=3, see \cite{batatia2024foundation} for the full list), Specifically hyperparameters for fine tuning are: learning rate 0.005, ema decay 0.995, batch size 8.
\subsection{Dealing with phonon imaginary modes}
Observing phonon imaginary modes for MOFs are common when the geometry optimization leads to a configuration which does not correspond to the minima of the potential energy surface. Among the curated dataset, NOTT-300 is a prototypical example that shows this phenomenon. We tested three methods to eliminate such spurious imaginary modes: i) Mode Mapping -- The process begins by identifying modes with negative frequencies and generating a structure by displacing the atoms along the imaginary mode. The structure is then geometry-optimized, followed by standard phonon calculations. Large enough displacements are needed to escape the local minima. This method is convenient for structures having only few imaginary modes. ii) Structure rattling -- this method is by far the most brute-force method, but surprisingly effective. Here, we used the ASE \cite{ase-paper,ISI:000175131400009} rattle implementation which perturbs atomic positions with random displacements drawn from a normal distribution of 0.01 standard deviation. It is possible to rattle the full structure or only atoms involved in the imaginary modes, such as the H atoms that hydroxylate the inorganic centre in NOTT-300. Once rattled, the structure is regularly geometry optimized, and iii) Molecular Dynamics -- in this method, we perform a low-temperature NVE ensemble simulation at T = 7.5 K or 10 K for as few as 40 steps, using either the MACE-MP-0b potential or the fine-tuned MACE-MP-MOF0 potential. We perform geometry optimization on the last frame of the resultant simulation. Importantly, specific atoms may be fixed in position if needed during the molecular dynamics simulation. All three methods mentioned above successfully eliminate imaginary modes, with methods ii) and iii) being preferred due to their ease of application and better suitability for handling structures with a large number of negative frequencies. The process of eliminating imaginary modes is stopped when the remaining negative frequencies become negligible ($\leq 10^{-4}$ THz).
\section{Data Availability}
All datasets that were used and/or generated in this work are publicly available in this repository: https://github.com/ddmms/data/tree/main/mace-mof-0
\section{Code Availability}
The scripts used to generate the training DFT dataset, the MACE-MP-MOF0 models are provided here: https://github.com/ddmms/data/tree/main/mace-mof-0
\section{Supplementary Information}
 The supplementary information is available \href{SI_revised.pdf}{here}
\section{Acknowledgments}
This research used resources of the National Energy Research Scientific Computing Center (NERSC), a Department of Energy Office of Science User Facility using NERSC award 'GenAI@NERSC'. PDK acknowledges financial support from U.S. National Science Foundation's "The Quantum Sensing Challenges for Transformational Advances in Quantum Systems (QuSeC-TAQS)" program. AME and FZ’s work
was also supported by Ada Lovelace Centre at STFC (https://adalovelacecentre.ac.uk/), Physical Sciences
Databases Infrastructure (https://psdi.ac.uk) under EPSRC grant no. EP/X032663/1  and EPSRC under grant no. EP/V028537/1.
A. S. R. acknowledges support via a Miller Research Fellowship from the Miller Institute for Basic Research in Science, University of California, Berkeley.
T. J. I. acknowledges support via the Bakar Institute of Digital Materials for the Planet (BIDMaP).

\bibliography{citation} %

\begin{thebibliography}{10}
\expandafter\ifx\csname url\endcsname\relax
  \def\url#1{\burl{#1}}\fi
\expandafter\ifx\csname urlprefix\endcsname\relax\def\urlprefix{URL }\fi
\providecommand{\bibinfo}[2]{#2}
\providecommand{\eprint}[2][]{\url{#2}}
\providecommand{\doi}[1]{\url{https://doi.org/#1}}
\bibcommenthead

\bibitem{mof1}
\bibinfo{author}{Furukawa, H.}, \bibinfo{author}{Cordova, K.~E.}, \bibinfo{author}{O’Keeffe, M.} \& \bibinfo{author}{Yaghi, O.~M.}
\newblock \bibinfo{title}{The chemistry and applications of metal-organic frameworks}.
\newblock \emph{\bibinfo{journal}{Science}} \textbf{\bibinfo{volume}{341}}, \bibinfo{pages}{1230444} (\bibinfo{year}{2013}).
\newblock \urlprefix\url{https://www.science.org/doi/abs/10.1126/science.1230444}.

\bibitem{mof2}
\bibinfo{author}{Yaghi, O.~M.} \emph{et~al.}
\newblock \bibinfo{title}{Reticular synthesis and the design of new materials}.
\newblock \emph{\bibinfo{journal}{Nature}} \textbf{\bibinfo{volume}{423}}, \bibinfo{pages}{705--714} (\bibinfo{year}{2003}).
\newblock \urlprefix\url{https://doi.org/10.1038/nature01650}.

\bibitem{mof3}
\bibinfo{author}{Li, H.}, \bibinfo{author}{Eddaoudi, M.}, \bibinfo{author}{O'Keeffe, M.} \& \bibinfo{author}{Yaghi, O.~M.}
\newblock \bibinfo{title}{Design and synthesis of an exceptionally stable and highly porous metal-organic framework}.
\newblock \emph{\bibinfo{journal}{Nature}} \textbf{\bibinfo{volume}{402}}, \bibinfo{pages}{276--279} (\bibinfo{year}{1999}).
\newblock \urlprefix\url{https://doi.org/10.1038/46248}.

\bibitem{water}
\bibinfo{author}{Hanikel, N.} \emph{et~al.}
\newblock \bibinfo{title}{Evolution of water structures in metal-organic frameworks for improved atmospheric water harvesting}.
\newblock \emph{\bibinfo{journal}{Science}} \textbf{\bibinfo{volume}{374}}, \bibinfo{pages}{454--459} (\bibinfo{year}{2021}).
\newblock \urlprefix\url{https://www.science.org/doi/abs/10.1126/science.abj0890}.

\bibitem{water2}
\bibinfo{author}{Zheng, Z.} \emph{et~al.}
\newblock \bibinfo{title}{High-yield, green and scalable methods for producing mof-303 for water harvesting from desert air}.
\newblock \emph{\bibinfo{journal}{Nature Protocols}} \textbf{\bibinfo{volume}{18}}, \bibinfo{pages}{136--156} (\bibinfo{year}{2023}).
\newblock \urlprefix\url{https://doi.org/10.1038/s41596-022-00756-w}.

\bibitem{D3TA02214E}
\bibinfo{author}{Hoffman, A. E.~J.} \emph{et~al.}
\newblock \bibinfo{title}{The role of phonons in switchable mofs: a model material perspective}.
\newblock \emph{\bibinfo{journal}{J. Mater. Chem. A}} \textbf{\bibinfo{volume}{11}}, \bibinfo{pages}{15286--15300} (\bibinfo{year}{2023}).
\newblock \urlprefix\url{http://dx.doi.org/10.1039/D3TA02214E}.

\bibitem{thermal}
\bibinfo{author}{Wieser, S.} \emph{et~al.}
\newblock \bibinfo{title}{Identifying the bottleneck for heat transport in metal–organic frameworks}.
\newblock \emph{\bibinfo{journal}{Advanced Theory and Simulations}} \textbf{\bibinfo{volume}{4}}, \bibinfo{pages}{2000211} (\bibinfo{year}{2021}).
\newblock \urlprefix\url{https://onlinelibrary.wiley.com/doi/abs/10.1002/adts.202000211}.

\bibitem{deform}
\bibinfo{author}{Formalik, F.}, \bibinfo{author}{Fischer, M.} \& \bibinfo{author}{Kuchta, B.}
\newblock \bibinfo{title}{Correlating phonons and deformations: A method for structural phase transformation analysis in metal--organic frameworks}.
\newblock \emph{\bibinfo{journal}{Crystal Growth \& Design}} \textbf{\bibinfo{volume}{23}}, \bibinfo{pages}{8962--8971} (\bibinfo{year}{2023}).
\newblock \urlprefix\url{https://doi.org/10.1021/acs.cgd.3c01013}.

\bibitem{supercond}
\bibinfo{author}{Takenaka, T.} \emph{et~al.}
\newblock \bibinfo{title}{Strongly correlated superconductivity in a copper-based metal-organic framework with a perfect kagome lattice}.
\newblock \emph{\bibinfo{journal}{Science Advances}} \textbf{\bibinfo{volume}{7}}, \bibinfo{pages}{eabf3996} (\bibinfo{year}{2021}).
\newblock \urlprefix\url{https://www.science.org/doi/abs/10.1126/sciadv.abf3996}.

\bibitem{phonons}
\bibinfo{author}{Kuchta, B.}, \bibinfo{author}{Formalik, F.}, \bibinfo{author}{Rogacka, J.}, \bibinfo{author}{Neimark, A.~V.} \& \bibinfo{author}{Firlej, L.}
\newblock \bibinfo{title}{Phonons in deformable microporous crystalline solids}.
\newblock \emph{\bibinfo{journal}{Zeitschrift für Kristallographie - Crystalline Materials}} \textbf{\bibinfo{volume}{234}}, \bibinfo{pages}{513--527} (\bibinfo{year}{2019}).
\newblock \urlprefix\url{https://doi.org/10.1515/zkri-2018-2152}.

\bibitem{PhysRev.136.B864}
\bibinfo{author}{Hohenberg, P.} \& \bibinfo{author}{Kohn, W.}
\newblock \bibinfo{title}{Inhomogeneous electron gas}.
\newblock \emph{\bibinfo{journal}{Phys. Rev.}} \textbf{\bibinfo{volume}{136}}, \bibinfo{pages}{B864--B871} (\bibinfo{year}{1964}).
\newblock \urlprefix\url{https://link.aps.org/doi/10.1103/PhysRev.136.B864}.

\bibitem{dft}
\bibinfo{author}{Kohn, W.}, \bibinfo{author}{Becke, A.~D.} \& \bibinfo{author}{Parr, R.~G.}
\newblock \bibinfo{title}{Density functional theory of electronic structure}.
\newblock \emph{\bibinfo{journal}{The Journal of Physical Chemistry}} \textbf{\bibinfo{volume}{100}}, \bibinfo{pages}{12974--12980} (\bibinfo{year}{1996}).
\newblock \urlprefix\url{https://doi.org/10.1021/jp960669l}.

\bibitem{dreizler_gross_dft_1990}
\bibinfo{author}{Dreizler, R.~M.} \& \bibinfo{author}{Gross, E. K.~U.}
\newblock \emph{\bibinfo{title}{Density Functional Theory: An Approach to the Quantum Many-Body Problem}} \bibinfo{edition}{1} edn.
\newblock Springer Book Archive (\bibinfo{publisher}{Springer-Verlag Berlin Heidelberg}, \bibinfo{address}{Berlin, Heidelberg}, \bibinfo{year}{1990}).
\newblock \bibinfo{note}{Published: 06 December 2012}.

\bibitem{PhysRevMaterials.3.116003}
\bibinfo{author}{Kamencek, T.}, \bibinfo{author}{Bedoya-Mart\'{\i}nez, N.} \& \bibinfo{author}{Zojer, E.}
\newblock \bibinfo{title}{Understanding phonon properties in isoreticular metal-organic frameworks from first principles}.
\newblock \emph{\bibinfo{journal}{Phys. Rev. Mater.}} \textbf{\bibinfo{volume}{3}}, \bibinfo{pages}{116003} (\bibinfo{year}{2019}).
\newblock \urlprefix\url{https://link.aps.org/doi/10.1103/PhysRevMaterials.3.116003}.

\bibitem{dftb3}
\bibinfo{author}{Gaus, M.}, \bibinfo{author}{Cui, Q.} \& \bibinfo{author}{Elstner, M.}
\newblock \bibinfo{title}{Dftb3: Extension of the self consistent charge density functional tight binding method (scc dftb)}.
\newblock \emph{\bibinfo{journal}{Journal of Chemical Theory and Computation}} \textbf{\bibinfo{volume}{7}}, \bibinfo{pages}{931--948} (\bibinfo{year}{2011}).
\newblock \urlprefix\url{https://doi.org/10.1021/ct100684s}.

\bibitem{tb}
\bibinfo{author}{Grimme, S.}, \bibinfo{author}{Bannwarth, C.} \& \bibinfo{author}{Shushkov, P.}
\newblock \bibinfo{title}{A robust and accurate tight-binding quantum chemical method for structures, vibrational frequencies, and noncovalent interactions of large molecular systems parametrized for all spd-block elements (z = 1--86)}.
\newblock \emph{\bibinfo{journal}{Journal of Chemical Theory and Computation}} \textbf{\bibinfo{volume}{13}}, \bibinfo{pages}{1989--2009} (\bibinfo{year}{2017}).
\newblock \urlprefix\url{https://doi.org/10.1021/acs.jctc.7b00118}.

\bibitem{D2CP00184E}
\bibinfo{author}{Nurhuda, M.}, \bibinfo{author}{Perry, C.~C.} \& \bibinfo{author}{Addicoat, M.~A.}
\newblock \bibinfo{title}{Performance of gfn1-xtb for periodic optimization of metal organic frameworks}.
\newblock \emph{\bibinfo{journal}{Phys. Chem. Chem. Phys.}} \textbf{\bibinfo{volume}{24}}, \bibinfo{pages}{10906--10914} (\bibinfo{year}{2022}).
\newblock \urlprefix\url{http://dx.doi.org/10.1039/D2CP00184E}.

\bibitem{core1}
\bibinfo{author}{Chung, Y.~G.} \emph{et~al.}
\newblock \bibinfo{title}{Advances, updates, and analytics for the computation-ready, experimental metal--organic framework database: Core mof 2019}.
\newblock \emph{\bibinfo{journal}{Journal of Chemical \& Engineering Data}} \textbf{\bibinfo{volume}{64}}, \bibinfo{pages}{5985--5998} (\bibinfo{year}{2019}).
\newblock \urlprefix\url{https://doi.org/10.1021/acs.jced.9b00835}.

\bibitem{core2}
\bibinfo{author}{Chung, Y.~G.} \emph{et~al.}
\newblock \bibinfo{title}{Computation-ready, experimental metal--organic frameworks: A tool to enable high-throughput screening of nanoporous crystals}.
\newblock \emph{\bibinfo{journal}{Chemistry of Materials}} \textbf{\bibinfo{volume}{26}}, \bibinfo{pages}{6185--6192} (\bibinfo{year}{2014}).
\newblock \urlprefix\url{https://doi.org/10.1021/cm502594j}.

\bibitem{uff}
\bibinfo{author}{Rappe, A.~K.}, \bibinfo{author}{Casewit, C.~J.}, \bibinfo{author}{Colwell, K.~S.}, \bibinfo{author}{Goddard, W. A.~I.} \& \bibinfo{author}{Skiff, W.~M.}
\newblock \bibinfo{title}{Uff, a full periodic table force field for molecular mechanics and molecular dynamics simulations}.
\newblock \emph{\bibinfo{journal}{Journal of the American Chemical Society}} \textbf{\bibinfo{volume}{114}}, \bibinfo{pages}{10024--10035} (\bibinfo{year}{1992}).
\newblock \urlprefix\url{https://doi.org/10.1021/ja00051a040}.

\bibitem{Brooks2009}
\bibinfo{author}{Brooks, B.~R.} \emph{et~al.}
\newblock \bibinfo{title}{{CHARMM: the biomolecular simulation program}}.
\newblock \emph{\bibinfo{journal}{Journal of Computational Chemistry}} \textbf{\bibinfo{volume}{30}}, \bibinfo{pages}{1545--1614} (\bibinfo{year}{2009}).

\bibitem{uff4mof}
\bibinfo{author}{Coupry, D.~E.}, \bibinfo{author}{Addicoat, M.~A.} \& \bibinfo{author}{Heine, T.}
\newblock \bibinfo{title}{Extension of the universal force field for metal--organic frameworks}.
\newblock \emph{\bibinfo{journal}{Journal of Chemical Theory and Computation}} \textbf{\bibinfo{volume}{12}}, \bibinfo{pages}{5215--5225} (\bibinfo{year}{2016}).
\newblock \urlprefix\url{https://doi.org/10.1021/acs.jctc.6b00664}.

\bibitem{Wieser2024}
\bibinfo{author}{Wieser, S.} \& \bibinfo{author}{Zojer, E.}
\newblock \bibinfo{title}{Machine learned force-fields for an ab-initio quality description of metal-organic frameworks}.
\newblock \emph{\bibinfo{journal}{npj Computational Materials}} \textbf{\bibinfo{volume}{10}}, \bibinfo{pages}{18} (\bibinfo{year}{2024}).
\newblock \urlprefix\url{https://doi.org/10.1038/s41524-024-01205-w}.

\bibitem{mofff}
\bibinfo{author}{Bureekaew, S.} \emph{et~al.}
\newblock \bibinfo{title}{Mof-ff – a flexible first-principles derived force field for metal-organic frameworks}.
\newblock \emph{\bibinfo{journal}{physica status solidi (b)}} \textbf{\bibinfo{volume}{250}}, \bibinfo{pages}{1128--1141} (\bibinfo{year}{2013}).
\newblock \urlprefix\url{https://onlinelibrary.wiley.com/doi/abs/10.1002/pssb.201248460}.

\bibitem{C6CP05106E}
\bibinfo{author}{Bristow, J.~K.}, \bibinfo{author}{Skelton, J.~M.}, \bibinfo{author}{Svane, K.~L.}, \bibinfo{author}{Walsh, A.} \& \bibinfo{author}{Gale, J.~D.}
\newblock \bibinfo{title}{A general forcefield for accurate phonon properties of metal–organic frameworks}.
\newblock \emph{\bibinfo{journal}{Phys. Chem. Chem. Phys.}} \textbf{\bibinfo{volume}{18}}, \bibinfo{pages}{29316--29329} (\bibinfo{year}{2016}).
\newblock \urlprefix\url{http://dx.doi.org/10.1039/C6CP05106E}.

\bibitem{nn}
\bibinfo{author}{Eckhoff, M.} \& \bibinfo{author}{Behler, J.}
\newblock \bibinfo{title}{From molecular fragments to the bulk: Development of a neural network potential for mof-5}.
\newblock \emph{\bibinfo{journal}{Journal of Chemical Theory and Computation}} \textbf{\bibinfo{volume}{15}}, \bibinfo{pages}{3793--3809} (\bibinfo{year}{2019}).
\newblock \urlprefix\url{https://doi.org/10.1021/acs.jctc.8b01288}.

\bibitem{Vandenhaute2023}
\bibinfo{author}{Vandenhaute, S.}, \bibinfo{author}{Cools-Ceuppens, M.}, \bibinfo{author}{DeKeyser, S.}, \bibinfo{author}{Verstraelen, T.} \& \bibinfo{author}{Van~Speybroeck, V.}
\newblock \bibinfo{title}{Machine learning potentials for metal-organic frameworks using an incremental learning approach}.
\newblock \emph{\bibinfo{journal}{npj Computational Materials}} \textbf{\bibinfo{volume}{9}}, \bibinfo{pages}{19} (\bibinfo{year}{2023}).
\newblock \urlprefix\url{https://doi.org/10.1038/s41524-023-00969-x}.

\bibitem{vasp}
\bibinfo{author}{Kresse, G.} \& \bibinfo{author}{Furthm\"uller, J.}
\newblock \bibinfo{title}{Efficient iterative schemes for ab initio total-energy calculations using a plane-wave basis set}.
\newblock \emph{\bibinfo{journal}{Phys. Rev. B}} \textbf{\bibinfo{volume}{54}}, \bibinfo{pages}{11169--11186} (\bibinfo{year}{1996}).
\newblock \urlprefix\url{https://link.aps.org/doi/10.1103/PhysRevB.54.11169}.

\bibitem{vasp1}
\bibinfo{author}{Kresse, G.} \& \bibinfo{author}{Joubert, D.}
\newblock \bibinfo{title}{From ultrasoft pseudopotentials to the projector augmented-wave method}.
\newblock \emph{\bibinfo{journal}{Phys. Rev. B}} \textbf{\bibinfo{volume}{59}}, \bibinfo{pages}{1758--1775} (\bibinfo{year}{1999}).
\newblock \urlprefix\url{https://link.aps.org/doi/10.1103/PhysRevB.59.1758}.

\bibitem{vasp2}
\bibinfo{author}{Kresse, G.} \& \bibinfo{author}{Furthmüller, J.}
\newblock \bibinfo{title}{Efficiency of ab-initio total energy calculations for metals and semiconductors using a plane-wave basis set}.
\newblock \emph{\bibinfo{journal}{Computational Materials Science}} \textbf{\bibinfo{volume}{6}}, \bibinfo{pages}{15--50} (\bibinfo{year}{1996}).
\newblock \urlprefix\url{https://www.sciencedirect.com/science/article/pii/0927025696000080}.

\bibitem{vasp3}
\bibinfo{author}{Kresse, G.} \& \bibinfo{author}{Hafner, J.}
\newblock \bibinfo{title}{Ab initio molecular dynamics for liquid metals}.
\newblock \emph{\bibinfo{journal}{Phys. Rev. B}} \textbf{\bibinfo{volume}{47}}, \bibinfo{pages}{558--561} (\bibinfo{year}{1993}).
\newblock \urlprefix\url{https://link.aps.org/doi/10.1103/PhysRevB.47.558}.

\bibitem{mtp}
\bibinfo{author}{Shapeev, A.~V.}
\newblock \bibinfo{title}{Moment tensor potentials: A class of systematically improvable interatomic potentials}.
\newblock \emph{\bibinfo{journal}{Multiscale Modeling \& Simulation}} \textbf{\bibinfo{volume}{14}}, \bibinfo{pages}{1153--1173} (\bibinfo{year}{2016}).
\newblock \urlprefix\url{https://doi.org/10.1137/15M1054183}.

\bibitem{wieser2023machine}
\bibinfo{author}{Wieser, S.} \& \bibinfo{author}{Zojer, E.}
\newblock \bibinfo{title}{Machine learned force-fields for an ab-initio quality description of metal-organic frameworks} (\bibinfo{year}{2023}).
\newblock \eprint{2308.01278}.

\bibitem{batatia2024foundation}
\bibinfo{author}{Batatia, I.} \emph{et~al.}
\newblock \bibinfo{title}{A foundation model for atomistic materials chemistry} (\bibinfo{year}{2024}).
\newblock \eprint{2401.00096}.

\bibitem{macearc}
\bibinfo{author}{Batatia, I.}, \bibinfo{author}{Kovács, D.~P.}, \bibinfo{author}{Simm, G. N.~C.}, \bibinfo{author}{Ortner, C.} \& \bibinfo{author}{Csányi, G.}
\newblock \bibinfo{title}{Mace: Higher order equivariant message passing neural networks for fast and accurate force fields} (\bibinfo{year}{2023}).
\newblock \urlprefix\url{https://arxiv.org/abs/2206.07697}.
\newblock \eprint{2206.07697}.

\bibitem{mptrj}
\bibinfo{author}{Deng, B.} \emph{et~al.}
\newblock \bibinfo{title}{Chgnet as a pretrained universal neural network potential for charge-informed atomistic modelling}.
\newblock \emph{\bibinfo{journal}{Nature Machine Intelligence}} \textbf{\bibinfo{volume}{5}}, \bibinfo{pages}{1031--1041} (\bibinfo{year}{2023}).
\newblock \urlprefix\url{https://doi.org/10.1038/s42256-023-00716-3}.

\bibitem{MOF}
\bibinfo{author}{Rosen, A.~S.} \emph{et~al.}
\newblock \bibinfo{title}{High-throughput predictions of metal–organic framework electronic properties: theoretical challenges, graph neural networks, and data exploration}.
\newblock \emph{\bibinfo{journal}{npj Comput Mater 8, 112}}  (\bibinfo{year}{2022}).

\bibitem{rosen2021machine}
\bibinfo{author}{Rosen, A.~S.} \emph{et~al.}
\newblock \bibinfo{title}{Machine learning the quantum-chemical properties of metal--organic frameworks for accelerated materials discovery}.
\newblock \emph{\bibinfo{journal}{Matter}} \textbf{\bibinfo{volume}{4}}, \bibinfo{pages}{1578--1597} (\bibinfo{year}{2021}).

\bibitem{csd}
\bibinfo{author}{Moghadam, P.~Z.} \emph{et~al.}
\newblock \bibinfo{title}{Development of a cambridge structural database subset: A collection of metal--organic frameworks for past, present, and future}.
\newblock \emph{\bibinfo{journal}{Chemistry of Materials}} \textbf{\bibinfo{volume}{29}}, \bibinfo{pages}{2618--2625} (\bibinfo{year}{2017}).
\newblock \urlprefix\url{https://doi.org/10.1021/acs.chemmater.7b00441}.

\bibitem{pormaker}
\bibinfo{author}{Lee, S.}, \bibinfo{author}{Kim, B.} \& \bibinfo{author}{Kim, J.}
\newblock \bibinfo{title}{Discovery of record-breaking metal-organic frameworks for methane storage using evolutionary algorithm and machine learning}.
\newblock \emph{\bibinfo{journal}{ChemRxiv}}  (\bibinfo{year}{2020}).
\newblock \bibinfo{note}{This content is a preprint and has not been peer-reviewed}.

\bibitem{ase-paper}
\bibinfo{author}{Larsen, A.~H.} \emph{et~al.}
\newblock \bibinfo{title}{The atomic simulation environment—a python library for working with atoms}.
\newblock \emph{\bibinfo{journal}{Journal of Physics: Condensed Matter}} \textbf{\bibinfo{volume}{29}}, \bibinfo{pages}{273002} (\bibinfo{year}{2017}).
\newblock \urlprefix\url{http://stacks.iop.org/0953-8984/29/i=27/a=273002}.

\bibitem{pymatgen}
\bibinfo{author}{Ong, S.~P.} \emph{et~al.}
\newblock \bibinfo{title}{Python materials genomics (pymatgen): A robust, open-source python library for materials analysis}.
\newblock \emph{\bibinfo{journal}{Computational Materials Science}} \textbf{\bibinfo{volume}{68}}, \bibinfo{pages}{314--hor}{Wang, Y.} \& \bibinfo{author}{Zhao, D.}
\newblock \bibinfo{title}{The chemistry and applications of hafnium and cerium(iv) metal–organic frameworks}.
\newblock \emph{\bibinfo{journal}{Chem. Soc. Rev.}} \textbf{\bibinfo{volume}{50}}, \bibinfo{pages}{4629--4683} (\bibinfo{year}{2021}).
\newblock \urlprefix\url{http://dx.doi.org/10.1039/D0CS00920B}.

\bibitem{harmonic}
\bibinfo{author}{Agne, M.~T.}, \bibinfo{author}{Anand, S.} \& \bibinfo{author}{Snyder, G.~J.}
\newblock \bibinfo{title}{Inherent anharmonicity of harmonic solids}.
\newblock \emph{\bibinfo{journal}{Research}} \textbf{\bibinfo{volume}{2022}}, \bibinfo{pages}{9786705} (\bibinfo{year}{2022}).
\newblock \urlprefix\url{https://doi.org/10.34133/2022/9786705}.

\bibitem{TOGO20151}
\bibinfo{author}{Togo, A.} \& \bibinfo{author}{Tanaka, I.}
\newblock \bibinfo{title}{First principles phonon calculations in materials science}.
\newblock \emph{\bibinfo{journal}{Scripta Materialia}} \textbf{\bibinfo{volume}{108}}, \bibinfo{pages}{1--5} (\bibinfo{year}{2015}).
\newblock \urlprefix\url{https://www.sciencedirect.com/science/article/pii/S1359646215003127}.

\bibitem{defect}
\bibinfo{author}{Hu, Z.}, \bibinfo{author}{Wang, Y.} \& \bibinfo{author}{Zhao, D.}
\newblock \bibinfo{title}{The chemistry and applications of hafnium and cerium(iv) metal–organic frameworks}.
\newblock \emph{\bibinfo{journal}{Chem. Soc. Rev.}} \textbf{\bibinfo{volume}{50}}, \bibinfo{pages}{4629--4683} (\bibinfo{year}{2021}).
\newblock \urlprefix\url{http://dx.doi.org/10.1039/D0CS00920B}.

\bibitem{Ziman2001}
\bibinfo{author}{Ziman, JM}
\newblock \emph{\bibinfo{title}{Electrons and Phonons: The Theory of Transport Phenomena in Solids}} Vol. \bibinfo{volume}{540} (\bibinfo{publisher}{Oxford University Press}, \bibinfo{year}{2001}).

\bibitem{mlff}
\bibinfo{author}{Ying, P.} \emph{et~al.}
\newblock \bibinfo{title}{Sub-micrometer phonon mean free paths in metal--organic frameworks revealed by machine learning molecular dynamics simulations}.
\newblock \emph{\bibinfo{journal}{ACS Applied Materials \& Interfaces}} \textbf{\bibinfo{volume}{15}}, \bibinfo{pages}{36412--36422} (\bibinfo{year}{2023}).
\newblock \urlprefix\url{https://doi.org/10.1021/acsami.3c07770}.

\bibitem{mlff}
\bibinfo{author}{Ying, P.} \emph{et~al.}
\newblock \bibinfo{title}{Sub-micrometer phonon mean free paths in metal--organic frameworks revealed by machine learning molecular dynamics simulations}.
\newblock \emph{\bibinfo{journal}{ACS Applied Materials \& Interfaces}} \textbf{\bibinfo{volume}{15}}, \bibinfo{pages}{36412--36422} (\bibinfo{year}{2023}).
\newblock \urlprefix\url{https://doi.org/10.1021/acsami.3c07770}.

\bibitem{dft-mof-74}
\bibinfo{author}{Kamencek, T.}, \bibinfo{author}{Schrode, B.}, \bibinfo{author}{Resel, R.}, \bibinfo{author}{Ricco, R.} \& \bibinfo{author}{Zojer, E.}
\newblock \bibinfo{title}{Understanding the origin of the particularly small and anisotropic thermal expansion of mof-74}.
\newblock \emph{\bibinfo{journal}{Advanced Theory and Simulations}} \textbf{\bibinfo{volume}{5}}, \bibinfo{pages}{2200031} (\bibinfo{year}{2022}).
\newblock \urlprefix\url{https://onlinelibrary.wiley.com/doi/abs/10.1002/adts.202200031}.

\bibitem{longrange}
\bibinfo{author}{Formalik, F.}, \bibinfo{author}{Neimark, A.~V.}, \bibinfo{author}{Rogacka, J.}, \bibinfo{author}{Firlej, L.} \& \bibinfo{author}{Kuchta, B.}
\newblock \bibinfo{title}{Pore opening and breathing transitions in metal-organic frameworks: Coupling adsorption and deformation}.
\newblock \emph{\bibinfo{journal}{Journal of Colloid and Interface Science}} \textbf{\bibinfo{volume}{578}}, \bibinfo{pages}{77--88} (\bibinfo{year}{2020}).
\newblock \urlprefix\url{https://www.sciencedirect.com/science/article/pii/S0021979720307165}.

\bibitem{opendac}
\bibinfo{author}{Sriram, A.} \emph{et al.}
\newblock \bibinfo{title}{The open dac 2023 dataset and challenges for sorbent discovery in direct air capture}.
\newblock \emph{\bibinfo{journal}{ACS Central Science}} \textbf{\bibinfo{volume}{10}}, \bibinfo{pages}{923--941} (\bibinfo{year}{2024}).
\newblock \urlprefix\url{https://doi.org/10.1021/acscentsci.3c01629}.

\bibitem{Shi2022}
\bibinfo{author}{Shi, Z.}, \bibinfo{author}{Weng, K.} \& \bibinfo{author}{Li, N.}
\newblock \bibinfo{title}{The atomic structure and mechanical properties of zif-4 under high pressure: Ab initio calculations}.
\newblock \emph{\bibinfo{journal}{Molecules (Basel, Switzerland)}} \textbf{\bibinfo{volume}{28}}, \bibinfo{pages}{22} (\bibinfo{year}{2022}).

\bibitem{irmof-14}
\bibinfo{author}{Yang, L.-M.}, \bibinfo{author}{Ravindran, P.}, \bibinfo{author}{Vajeeston, P.} \& \bibinfo{author}{Tilset, M.}
\newblock \bibinfo{title}{Properties of irmof-14 and its analogues m-irmof-14 (m = cd{,} alkaline earth metals): electronic structure{,} structural stability{,} chemical bonding{,} and optical properties}.
\newblock \emph{\bibinfo{journal}{Phys. Chem. Chem. Phys.}} \textbf{\bibinfo{volume}{14}}, \bibinfo{pages}{4713--4723} (\bibinfo{year}{2012}).
\newblock \urlprefix\url{http://dx.doi.org/10.1039/C2CP24091B"}.

\bibitem{hoffman53}
\bibinfo{author}{Hoffman, A.~E.}, \bibinfo{author}{Wieme, J.}, \bibinfo{author}{Rogge, S.~M.}, \bibinfo{author}{Vanduyfhuys, L.} \& \bibinfo{author}{Speybroeck, V.~V.}
\newblock \bibinfo{title}{The impact of lattice vibrations on the macroscopic breathing behavior of mil-53(al)}.
\newblock \emph{\bibinfo{journal}{Zeitschrift für Kristallographie - Crystalline Materials}} \textbf{\bibinfo{volume}{234}}, \bibinfo{pages}{529--545} (\bibinfo{year}{2019}).
\newblock \urlprefix\url{https://doi.org/10.1515/zkri-2018-2154}.

\bibitem{zif8}
\bibinfo{author}{Dissegna, S.} \emph{et~al.}
\newblock \bibinfo{title}{Tuning the mechanical response of metal--organic frameworks by defect engineering}.
\newblock \emph{\bibinfo{journal}{Journal of the American Chemical Society}} \textbf{\bibinfo{volume}{140}}, \bibinfo{pages}{11581--11584} (\bibinfo{year}{2018}).
\newblock \urlprefix\url{https://doi.org/10.1021/jacs.8b07098}.

\bibitem{mil-125}
\bibinfo{author}{Yot, P.~G.} \emph{et~al.}
\newblock \bibinfo{title}{Exploration of the mechanical behavior of metal organic frameworks uio-66(zr) and mil-125(ti) and their nh2 functionalized versions}.
\newblock \emph{\bibinfo{journal}{Dalton Trans.}} \textbf{\bibinfo{volume}{45}}, \bibinfo{pages}{4283--4288} (\bibinfo{year}{2016}).
\newblock \urlprefix\url{http://dx.doi.org/10.1039/C5DT03621F}.

\bibitem{mil-53}
\bibinfo{author}{Yot, P.~G.} \emph{et~al.}
\newblock \bibinfo{title}{Metal–organic frameworks as potential shock absorbers: the case of the highly flexible mil-53(al)}.
\newblock \emph{\bibinfo{journal}{Chem. Commun.}} \textbf{\bibinfo{volume}{50}}, \bibinfo{pages}{9462--9464} (\bibinfo{year}{2014}).
\newblock \urlprefix\url{http://dx.doi.org/10.1039/C4CC03853C}.

\bibitem{uio66-ce}
\bibinfo{author}{Redfern, L.~R.} \emph{et~al.}
\newblock \bibinfo{title}{Isolating the role of the node-linker bond in the compression of uio-66 metal--organic frameworks}.
\newblock \emph{\bibinfo{journal}{Chemistry of Materials}} \textbf{\bibinfo{volume}{32}}, \bibinfo{pages}{5864--5871} (\bibinfo{year}{2020}).
\newblock \urlprefix\url{https://doi.org/10.1021/acs.chemmater.0c01922}.

\bibitem{uio67}
\bibinfo{author}{Redfern, L.~R.} \emph{et~al.}
\newblock \bibinfo{title}{Porosity dependence of compression and lattice rigidity in metal--organic framework series}.
\newblock \emph{\bibinfo{journal}{Journal of the American Chemical Society}} \textbf{\bibinfo{volume}{141}}, \bibinfo{pages}{4365--4371} (\bibinfo{year}{2019}).
\newblock \urlprefix\url{https://doi.org/10.1021/jacs.8b13009}.

\bibitem{C5CP01307K}
\bibinfo{author}{Cliffe, M.~J.}, \bibinfo{author}{Hill, J.~A.}, \bibinfo{author}{Murray, C.~A.}, \bibinfo{author}{Coudert, F.-X.} \& \bibinfo{author}{Goodwin, A.~L.}
\newblock \bibinfo{title}{Defect-dependent colossal negative thermal expansion in uio-66(hf) metal–organic framework}.
\newblock \emph{\bibinfo{journal}{Phys. Chem. Chem. Phys.}} \textbf{\bibinfo{volume}{17}}, \bibinfo{pages}{11586--11592} (\bibinfo{year}{2015}).
\newblock \urlprefix\url{http://dx.doi.org/10.1039/C5CP01307K}.

\bibitem{dft-66}
\bibinfo{author}{Vornholt, S.~M.}, \bibinfo{author}{Chen, Z.}, \bibinfo{author}{Hofmann, J.} \& \bibinfo{author}{Chapman, K.~W.}
\newblock \bibinfo{title}{Node distortions in uio-66 inform negative thermal expansion mechanisms: Kinetic effects, frustration, and lattice hysteresis}.
\newblock \emph{\bibinfo{journal}{Journal of the American Chemical Society}} \textbf{\bibinfo{volume}{146}}, \bibinfo{pages}{16977--16981} (\bibinfo{year}{2024}).
\newblock \urlprefix\url{https://doi.org/10.1021/jacs.4c05313}.

\bibitem{nte-dft}
\bibinfo{author}{Wang, L.} \emph{et~al.}
\newblock \bibinfo{title}{Large negative thermal expansion provided by metal--organic framework mof--5: A first--principles study}.
\newblock \emph{\bibinfo{journal}{Materials Chemistry and Physics}} \textbf{\bibinfo{volume}{175}}, \bibinfo{pages}{138--145} (\bibinfo{year}{2016}).
\newblock \urlprefix\url{https://www.sciencedirect.com/science/article/pii/S0254058416301511}.

\bibitem{nte-md}
\bibinfo{author}{Han, S.~S.} \& \bibinfo{author}{Goddard, W.~A.}
\newblock \bibinfo{title}{Metal organic frameworks provide large negative thermal expansion behavior}.
\newblock \emph{\bibinfo{journal}{The Journal of Physical Chemistry C}} \textbf{\bibinfo{volume}{111}}, \bibinfo{pages}{15185--15191} (\bibinfo{year}{2007}).
\newblock \urlprefix\url{https://doi.org/10.1021/jp075389s}.

\bibitem{nte-exp}
\bibinfo{author}{Lock, N.} \emph{et~al.}
\newblock \bibinfo{title}{Elucidating negative thermal expansion in mof--5}.
\newblock \emph{\bibinfo{journal}{The Journal of Physical Chemistry C}} \textbf{\bibinfo{volume}{114}}, \bibinfo{pages}{16181--16186} (\bibinfo{year}{2010}).
\newblock \urlprefix\url{https://doi.org/10.1021/jp103212z}.

\bibitem{cof}
\bibinfo{author}{Lyle, S.~J.}, \bibinfo{author}{Waller, P.~J.} \& \bibinfo{author}{Yaghi, O.~M.}
\newblock \bibinfo{title}{Covalent organic frameworks: Organic chemistry extended into two and three dimensions}.
\newblock \emph{\bibinfo{journal}{Trends in Chemistry}} \textbf{\bibinfo{volume}{1}}, \bibinfo{pages}{172--184} (\bibinfo{year}{2019}).
\newblock \urlprefix\url{https://www.sciencedirect.com/science/article/pii/S2589597419300310}.
\newblock \bibinfo{note}{Special Issue Part Two: Big Questions in Chemistry}.

\bibitem{atomate2}
\bibinfo{author}{Ganose, A.} \emph{et~al.}
\newblock \bibinfo{title}{atomate2}.
\newblock \urlprefix\url{https://github.com/materialsproject/atomate2}.

\bibitem{pbegga}
\bibinfo{author}{Perdew, J.~P.}, \bibinfo{author}{Burke, K.} \& \bibinfo{author}{Ernzerhof, M.}
\newblock \bibinfo{title}{Generalized gradient approximation made simple}.
\newblock \emph{\bibinfo{journal}{Phys. Rev. Lett.}} \textbf{\bibinfo{volume}{77}}, \bibinfo{pages}{3865--3868} (\bibinfo{year}{1996}).
\newblock \urlprefix\url{https://link.aps.org/doi/10.1103/PhysRevLett.77.3865}.

\bibitem{paw}
\bibinfo{author}{Bl\"ochl, P.~E.}
\newblock \bibinfo{title}{Projector augmented-wave method}.
\newblock \emph{\bibinfo{journal}{Phys. Rev. B}} \textbf{\bibinfo{volume}{50}}, \bibinfo{pages}{17953--17979} (\bibinfo{year}{1994}).
\newblock \urlprefix\url{https://link.aps.org/doi/10.1103/PhysRevB.50.17953}.

\bibitem{bj}
\bibinfo{author}{Grimme, S.}, \bibinfo{author}{Ehrlich, S.} \& \bibinfo{author}{Goerigk, L.}
\newblock \bibinfo{title}{Effect of the damping function in dispersion corrected density functional theory}.
\newblock \emph{\bibinfo{journal}{Journal of Computational Chemistry}} \textbf{\bibinfo{volume}{32}}, \bibinfo{pages}{1456--1465} (\bibinfo{year}{2011}).
\newblock \urlprefix\url{https://onlinelibrary.wiley.com/doi/abs/10.1002/jcc.21759}.

\bibitem{d3}
\bibinfo{author}{Grimme, S.}, \bibinfo{author}{Antony, J.}, \bibinfo{author}{Ehrlich, S.} \& \bibinfo{author}{Krieg, H.}
\newblock \bibinfo{title}{{A consistent and accurate ab initio parametrization of density functional dispersion correction (DFT-D) for the 94 elements H-Pu}}.
\newblock \emph{\bibinfo{journal}{The Journal of Chemical Physics}} \textbf{\bibinfo{volume}{132}}, \bibinfo{pages}{154104} (\bibinfo{year}{2010}).
\newblock \urlprefix\url{https://doi.org/10.1063/1.3382344}.

\bibitem{func}
\bibinfo{author}{Nazarian, D.}, \bibinfo{author}{Ganesh, P.} \& \bibinfo{author}{Sholl, D.~S.}
\newblock \bibinfo{title}{Benchmarking density functional theory predictions of framework structures and properties in a chemically diverse test set of metal–organic frameworks}.
\newblock \emph{\bibinfo{journal}{J. Mater. Chem. A}} \textbf{\bibinfo{volume}{3}}, \bibinfo{pages}{22432--22440} (\bibinfo{year}{2015}).
\newblock \urlprefix\url{http://dx.doi.org/10.1039/C5TA03864B}.

\bibitem{ISI:000175131400009}
\bibinfo{author}{Bahn, S.~R.} \& \bibinfo{author}{Jacobsen, K.~W.}
\newblock \bibinfo{title}{An object-oriented scripting interface to a legacy electronic structure code}.
\newblock \emph{\bibinfo{journal}{Comput. Sci. Eng.}} \textbf{\bibinfo{volume}{4}}, \bibinfo{pages}{56--66} (\bibinfo{year}{2002}).

\end{thebibliography}

\end{document}